\documentclass[lettersize,journal]{IEEEtran}
\usepackage{amsmath,amsfonts}
\usepackage{amssymb}
\usepackage{amsthm} 
\usepackage{array}
\usepackage{cite}
\usepackage[caption=false,font=normalsize,labelfont=sf,textfont=sf]{subfig}
\usepackage[colorlinks,urlcolor=blue,linkcolor=blue,citecolor=blue]{hyperref}
\usepackage{textcomp}
\usepackage{stfloats}
\usepackage{url}
\usepackage{verbatim}
\usepackage{graphicx}
\usepackage{xcolor}
\usepackage{algorithm}
\usepackage{enumitem}
\usepackage{svg}
\usepackage[parent]{overpic}
\usepackage{flushend}
\usepackage{braket}
\usepackage{bbm}
\usepackage{algpseudocode} 
\newtheorem{lemma}{Lemma}  
\newtheorem{proposition}{Proposition}

\def\BibTeX{{\rm B\kern-.05em{\sc i\kern-.025em b}\kern-.08em
    T\kern-.1667em\lower.7ex\hbox{E}\kern-.125emX}}
\usepackage{balance}

\begin{document}
	
	\title{Entanglement  Allocation for Quantum Integrated Sensing and Communication}
    
	 \author{Shiva Kazemi Taskooh, Mehdi Monemi, Ekram Hossain,  Georges Kaddoum, and Mehdi Rasti
   \thanks{ S. Kazemi Taskooh and E. Hossain are with the Department of Electrical and Computer Engineering, University of Manitoba, Winnipeg, Manitoba, Canada (e-mails: \{shiva.kazemitaskooh, ekram.hossain\}@umanitoba.ca). \\
   M. Monemi and M. Rasti are with the Centre for Wireless Communication, University of Oulu, Oulu, Finland (emails: \{mehdi.monemi, mehdi.rasti\}@oulu.fi). \\
   G. Kaddoum is with the Department of Electrical Engineering,  \'{E}cole de technologie sup\'{e}rieure, Montr\'{e}al, Qu\'{e}bec, Canada (e-mail: georges.kaddoum@etsmtl.ca).
   }
 }
\maketitle	
	
\begin{abstract}
     Jointly supporting communication and distributed sensing over a shared, limited pool of entanglement resources is an attractive prospect for future quantum networks. We develop a network-level quantum integrated sensing and communication (Q-ISAC) framework in which a quantum base station dynamically allocates elementary Bell pairs between communication users and distributed sensing. Sensing resources are adaptively partitioned among multiple GHZ probes with dynamically selected sizes and sensing-node assignments. We derive an effective quantum Fisher information model accounting for probabilistic fusion, transmission loss, and memory decoherence, and formulate a finite-horizon allocation problem subject to communication QoS, expected sensing-information, sensing-reliability, and hardware constraints. A hierarchical constrained proximal policy optimization (HC-PPO) framework combines learned discrete sensing allocation with analytical continuous communication allocation. 
    Numerical results show that GHZ probing requires approximately $27$--$30\%$ fewer sensing qubits than separable probing under modest probe-size caps and $65$--$67\%$ fewer at the highest considered target under the same sensing requirements. Adaptive multi-GHZ partitioning provides greater robustness than single-GHZ sensing, while adaptive communication--sensing sharing outperforms the evaluated fixed splits. The proposed HC-PPO framework satisfies the joint requirements in at least $99.9\%$ of test instances across the considered sensing targets and fusion probabilities. Monte Carlo evaluation separately shows that realized terminal information reached its target in $96$--$98\%$ of realizations, distinguishing expected-information satisfaction from realized sensing performance.
\end{abstract}


\section{Introduction}\label{introduction}
\subsection{Background}\label{background}
    Quantum networks extend conventional communication networks by interconnecting spatially separated quantum devices through quantum channels and shared entanglement \cite{Shiva-tutorial}. These networks are expected to support a variety of applications, including secure quantum communication and quantum sensing. In these applications, entanglement constitutes a fundamental network resource whose generation, distribution, storage, and consumption determine which services  can be provided to remote quantum nodes. However, establishing high-quality entanglement over distance remains a central challenge because optical transmission is inherently lossy and entanglement-generation operations are typically probabilistic \cite{Shiva-tutorial}.

    Quantum communication uses quantum states and correlations to enable capabilities such as secure information exchange, teleportation, and entanglement-assisted information transmission \cite{Shiva-tutorial}.
    On the other hand, quantum sensing employs quantum resources such as entanglement to estimate physical parameters with a precision surpassing classical sensing limits. When multiple spatially separated sensors are involved, distributed quantum sensing can further exploit correlations among networked probes to estimate global or spatially distributed parameters. Performance of such sensing systems is commonly characterized through the quantum Fisher information (QFI), which sets the quantum Cramér--Rao lower bound on the achievable estimation variance \cite{global-phase}.

    Future quantum networks are expected to not only exchange information among remote nodes, but also to acquire information about the physical environment where those nodes operate. 
    Supporting these functions through completely independent infrastructures can require separate quantum-state sources, transmission resources, and processing capabilities, which may result in inefficient use of scarce quantum resources. Moreover, quantum communication and quantum sensing share important physical operations, as both rely on the preparation, transmission, and measurement of quantum states, frequently exploiting entanglement as a common resource. These commonalities make the coexistence of communication and sensing particularly natural in quantum networks and motivate quantum integrated sensing and communication  (Q-ISAC) \cite{qisac-monemi,QISAC-entanglement}. 
    Q-ISAC enables communication and sensing to share quantum hardware, transmission links, and quantum resources. Such integration can improve resource utilization, reduce duplicated infrastructure, and enable flexible balancing of the available resources between communication performance and sensing accuracy.

    Entanglement plays a central role in this integrated architecture. For quantum communication, bipartite entanglement in the form of Bell pairs provides a basic resource for tasks such as teleportation, superdense coding, and entanglement-based networking \cite{Shiva-tutorial}. For distributed sensing, multipartite entangled states can correlate spatially separated sensing nodes and enhance their collective sensitivity to a parameter of interest. In particular, Greenberger--Horne--Zeilinger (GHZ) states are a canonical resource for distributed phase sensing, as the phase information accumulated by multiple sensing qubits can be coherently combined.  Under ideal phase-estimation conditions, GHZ-based entangled probes can achieve a higher QFI than corresponding separable probes and approach Heisenberg-limited precision \cite{global-phase}. Multipartite  entanglement can also be constructed from elementary bipartite resources; for instance, Bell pairs can be combined through fusion operations to generate GHZ states shared among several nodes \cite{GHZ-fusion}. Consequently, elementary entanglement can be viewed as a common resource that may be  directly consumed by communication users or transformed into multipartite probes for sensing.

    However, jointly supporting these two services introduces several coupled resource-management challenges. First, entanglement generation capacity is inherently limited and frequently probabilistic; therefore, allocating additional Bell pairs to one functionality necessarily reduces the resources available to the other \cite{QISAC-entanglement}. Second, communication and sensing consume entanglement in fundamentally different forms: while communication may directly use bipartite Bell pairs, distributed sensing may require the preparation of multipartite states through additional fusion operations. Third, whereas practical quantum links are affected by photon loss, quantum memories and multipartite states are vulnerable to decoherence and other imperfections. These effects are particularly important for large entangled probes, whose ideal metrological advantage can rapidly deteriorate in realistic noisy environments \cite{GHZ-partitioning}. Finite state generation, fusion, quantum memory, and processing capabilities further restrict the number and structure of quantum states that can be generated and distributed \cite{avis-analysis}.
    
    These considerations make Q-ISAC resource management intrinsically coupled across communication and sensing. While a resource allocation favoring communication can increase the delivered entanglement rate, it may leave insufficient resources for the required sensing accuracy, whereas allocating too many resources to multipartite sensing can degrade communication quality of service (QoS).  Therefore, effective Q-ISAC operation requires jointly accounting for communication requirements, sensing accuracy, entanglement generation and distribution, and physical limitations of the underlying quantum network.

\subsection{Related Works}\label{related-work}
    Previous research related to Q-ISAC spans two interconnected directions: entanglement-assisted distributed sensing \cite{global-phase,QFI-constant,quantum-advantage,GHZ-partitioning} and Q-ISAC \cite{wang-joint,QISAC-entanglement,variational,qisac-variational,isac-radar}.

    In entanglement-assisted distributed sensing, \cite{global-phase} established the metrological advantage of multipartite entanglement for estimating global functions of locally encoded parameters and showed that GHZ-based protocols can attain the corresponding generalized Heisenberg limit. Furthermore, in \cite{QFI-constant}, the authors characterized the conditions under which entanglement improves multiparameter estimation in networked quantum sensors, showing that its advantage depends on the structure of the estimation task. Moving towards practical network conditions, \cite{quantum-advantage} analyzed distributed sensing with GHZ-diagonal probes under imperfect entanglement generation, local-operation errors, and dephasing, and identified conditions under which a quantum sensing advantage can be preserved. More recently, \cite{GHZ-partitioning} showed that partitioning a sensing network into multiple smaller GHZ states can improve QFI relative to a single global GHZ state under state-preparation errors, sensor loss, and dephasing. However, while the studies reviewed above established the benefits and limitations of multipartite entanglement for distributed sensing, they did not consider the dynamic sharing of entanglement resources between communication and sensing services. 
    In the Q-ISAC direction, \cite{wang-joint} provided an information-theoretic characterization of joint quantum communication and sensing over a classical--quantum channel through a rate--detection-error-exponent tradeoff. In another relevant study \cite{QISAC-entanglement}, the authors developed an entanglement-based framework for secure communication and remote phase sensing and investigated their interplay under limited entanglement resources. Learning-based designs were considered in \cite{variational}, which combined variational quantum sensing with anonymous quantum communication, and in \cite{qisac-variational}, which employed parameterized quantum circuits and classical neural networks to jointly optimize superdense-coding-based communication and parameter estimation. In a different architecture, \cite{isac-radar} integrated quantum-illumination radar with classical full-duplex communication and optimized transmit power and beamforming subject to sensing requirements.

\subsection{Motivation, Contributions, and Structure}\label{motivation}
    The studies reviewed above established the benefits of multipartite entanglement for distributed sensing and demonstrated several Q-ISAC architectures in quantum networks. However, the network-level co-design of quantum communication, distributed sensing, and multipartite-entanglement resources remains largely unexplored. Such co-design is important because communication and sensing compete for the same entanglement-generation and quantum-hardware resources, so optimizing them independently can lead to inefficient resource use or failure to satisfy one service while favoring the other. Moreover, the appropriate sensing-entanglement structure depends on channel loss, fusion reliability, decoherence, and the required sensing accuracy. Seeking to address this gap, in this study, we develop a dynamic Q-ISAC framework that jointly adapts the communication--sensing resource sharing and the multipartite sensing configuration over a common Bell-pair budget. The main contributions can be summarized as follows:    
    \begin{itemize}[leftmargin=*]
    
        \item We develop a network-level Q-ISAC model that jointly allocates a shared supply of elementary Bell pairs to quantum communication and distributed multipartite sensing.
    
        \item We introduce adaptive multi-GHZ sensing with dynamically selected probe numbers, sizes, and sensing-node assignments. We also derive an effective QFI model accounting for probabilistic Bell-pair fusion, transmission loss, and quantum-memory decoherence.
    
        \item We jointly adapt communication--sensing resource sharing and multipartite sensing configurations over a finite horizon, subject to communication QoS, expected sensing-information and reliability requirements, and finite Bell-pair, processing, memory, and fusion resources.
    
         \item To realize this joint adaptation, we develop a hierarchical constrained proximal policy optimization (HC-PPO) framework combining feasibility-aware learned sensing allocation with analytical communication allocation. Results demonstrate sensing-resource savings over separable probing, greater robustness of multi-GHZ partitioning than single-GHZ sensing, and improved performance from adaptive communication--sensing sharing over fixed splitting.
    
    \end{itemize} 
    The remainder of this paper is organized as follows. Section \ref{system model} presents the system model, Section \ref{problem statement} formulates the resource-allocation problem, and Section \ref{algorithm} develops the proposed HC-PPO solution. Sections \ref{simulation} and \ref{conclusion} report the simulation results and conclude the paper, respectively.
    
\section{System Model and Notations}\label{system model}
    \begin{figure}
    \begin{center}
        \begin{overpic}[width=0.775\linewidth]{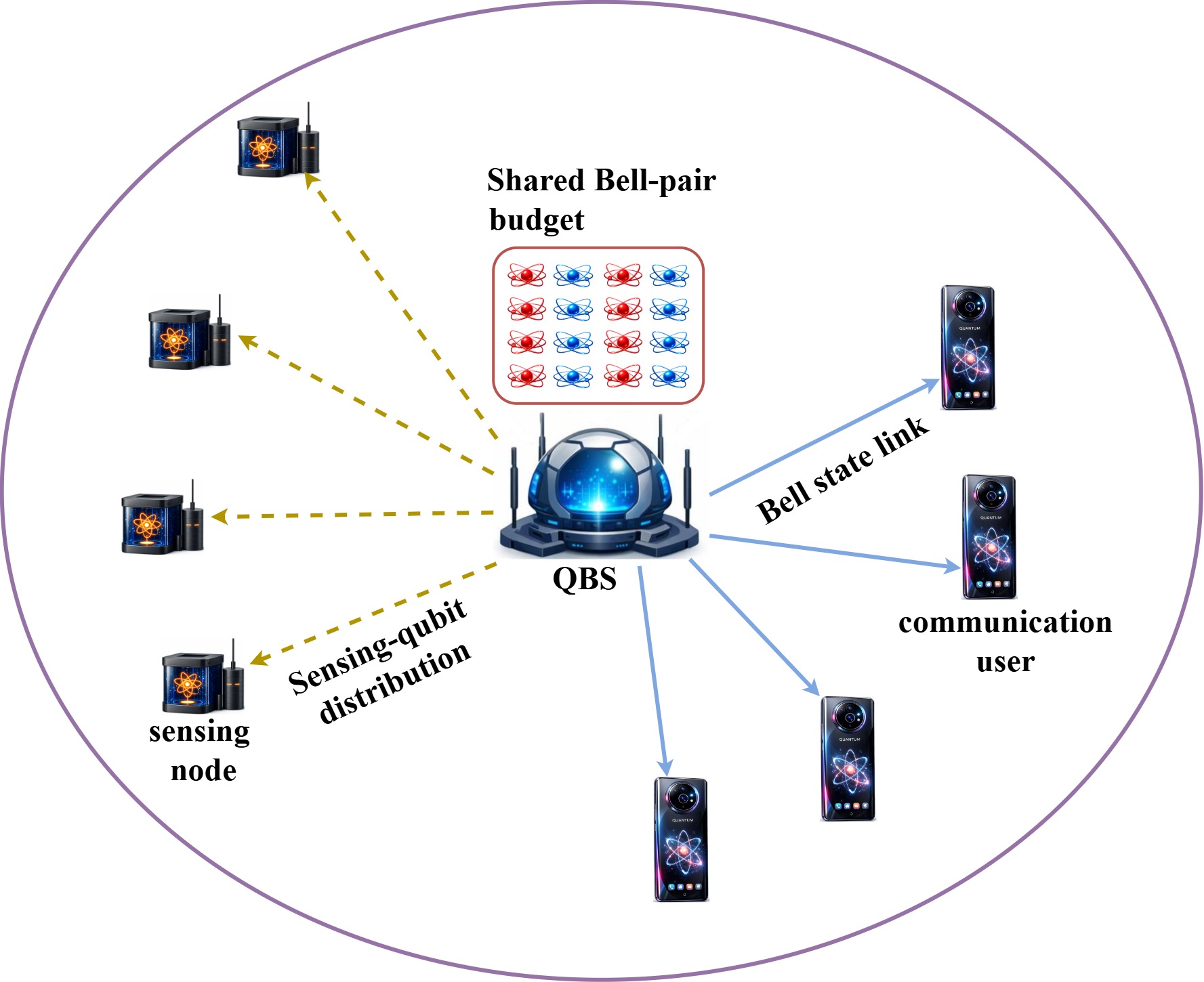}
        \put (26,72.5) {\small$\displaystyle n_{1,g,t}$}
            \put (14,58.5) {\small$\displaystyle n_{2,g,t}$}
            \put (10,42.5) {\small$\displaystyle \vdots$}
            \put (10.5,29.5) {\small$\displaystyle n_{s,g,t}$}
            \put (50,62) {\tiny$\displaystyle N^\mathrm{max}$}
        \end{overpic}
        \caption{Q-ISAC system showing one representative GHZ probe;    multiple independent probes may coexist in each slot.}
        \label{fig:systemmodel}
    \end{center}
    \end{figure}

    As illustrated in Fig.~\ref{fig:systemmodel}, we consider a Q-ISAC network, where a quantum base station (QBS) simultaneously serves a set of quantum communication users $\mathcal{U}=\{1,2,\ldots,U\}$ and a set of sensing nodes $\mathcal{S}=\{1,2,\ldots,S\}$ via optical channels including free-space optical (FSO) and fiber links. The system operates over a finite horizon $\mathcal T=\{1,2,\ldots,T\}$, where each slot has duration $\tau$. Channel transmissivities are assumed to remain constant within each optimization horizon, but may vary between different horizons.
    
\subsection{Channel Model}\label{channels}

    To make the analysis more general, we consider Q-ISAC over both FSO and fiber-optic links. Although their loss mechanisms differ, both can be  modeled as bosonic pure-loss channels with transmissivity $\eta_i^\zeta\in[0,1]$, representing the single-photon transmission probability \cite{bosonic}. Transmissivity accounts for medium-dependent losses, such as  fiber attenuation or FSO path loss and atmospheric effects, allowing a common analysis of both channels. Here, $\zeta\in\{\mathrm{sen},\mathrm{com}\}$ denotes sensing and communication links, respectively.
    
\subsubsection{FSO Channels} 
    Transmission through an FSO channel is affected by atmospheric attenuation, atmospheric turbulence, and pointing errors. The overall channel transmissivity between each node $i\in\mathcal{U} \cup \mathcal{S}$ (either communication users or sensing nodes) and the QBS is modeled as follows:
    \begin{equation}\label{gain-FSO}
        \eta_{i}^{{\zeta}}= h^{\mathrm{atm}}_{i}\, h^{\mathrm{turb}}_{i}\,
                   h^{\mathrm{pe}}_{i}, \quad \forall {\zeta} \in \{\mathrm{sen},\mathrm{com}\},
        \quad \forall i\in\mathcal{U} \cup \mathcal{S}.
    \end{equation} 
    
    Atmospheric attenuation $h^{\mathrm{atm}}_{i} = 10^{-\kappa d_{i}/10}$ accounts for absorption and scattering, where $\kappa$ is the weather-dependent attenuation coefficient, while $d_{i}$ is the distance between communication user or sensing node $i$ to the QBS.

    Under moderate-to-strong atmospheric turbulence, the turbulence-induced irradiance gain can be modeled using a  Gamma--Gamma distribution \cite{channel-modeling,chehimi-ris}.  In this study, we focus on resource allocation under average channel conditions. Therefore, following \cite{channel-modeling}, the turbulence gain is replaced by its normalized mean as $h_i^{\mathrm{turb}}  \approx\mathbb{E}\! \left[h_i^{\mathrm{turb}}\right] =1$\footnote{Instantaneous turbulence realizations are left for future work.}.

    Pointing errors in terrestrial FSO links arise from mechanical jitter and misalignment~\cite{chehimi-ris}. For resource-allocation design, pointing loss is approximated by evaluating aperture coupling at a representative displacement $d_i\sigma_\theta$ as  $h^{\mathrm{pe}}_{i} \approx A_{i}^o
    \exp\!\left(-\frac{2 d_{i}^2 \sigma_\theta^2}{W_{i}^2}\right)$,  
    where $\sigma_\theta^2$ is the angular jitter variance, $W_{i}^2 = w_{i}^2\sqrt{\pi}\,\mathrm{erf}(v_{i})/            (2v_{i}e^{-v_{i}^2})$ is the equivalent beam width, $w_{i}=\phi_d d_{i}$,      $v_{i}=\sqrt{\pi}r_a/(\sqrt{2}w_{i})$, $r_a$ is the receiver aperture radius, $A_{i}^o=[\mathrm{erf}(v_{i})]^2$, and, finally, $\phi_d$ is the beam divergence angle \cite{chehimi-ris,pointing-error}. Of note, this deterministic approximation captures average misalignment loss for resource-allocation design, instead of the full instantaneous pointing-error distribution.

    \subsubsection{Optical Fiber Links} 
    Transmission over optical fiber links is primarily affected by propagation loss due to absorption and Rayleigh scattering in the fiber medium. Unlike FSO channels, fiber‑optic links are largely immune to atmospheric turbulence and pointing errors; therefore, their channel transmissivity mainly depends on the fiber attenuation and link length. Channel transmissivity between communication user/sensing node $i$ and the QBS over an optical fiber link is modeled as follows \cite{fiber-link}:
    \begin{equation}\label{gain-fiber}
        \begin{aligned}
            \eta_i^\zeta = 10^{-\kappa_{\mathrm{f}} d_i / 10}, \quad \forall {\zeta} \in \{\mathrm{sen},\mathrm{com}\},
        \quad \forall i\in\mathcal{U} \cup \mathcal{S},
        \end{aligned}
    \end{equation}
    where $\kappa_{\mathrm{f}}$ is the fiber attenuation coefficient in $\mathrm{dB}/\mathrm{km}$. 
    
\subsection{Communication Model}
    It is assumed that the QBS is equipped with an entangled-pair source providing at most $N^{\max}$ elementary Bell pairs per slot of duration $\tau$. This budget is shared between communication users and distributed sensing.
    
    For communication, the QBS retains one qubit of each generated Bell pair in its local quantum memory and transmits the other qubit to the intended communication user through an optical link. At this stage, the local operations and QBS-side storage are assumed ideal, while the transmission loss is captured through the pure-loss channel transmissivity $\eta_i^{\mathrm{com}}$. 
    
    Let $r_{i,t}\geq 0$ denote the Bell-pair generation rate, measured in Bell pairs per unit time, allocated to communication user $i\in\mathcal U$ during slot $t\in\mathcal T$. Hence, $\tau r_{i,t}$ elementary Bell pairs are allocated to user $i$ during slot $t$. The delivered entanglement rate to user $i$ is computed as follows: 
    \begin{equation}
    \label{EDR}
        R_{i,t}
        =
        r_{i,t}\eta_i^{\mathrm{com}},
        \qquad
        i\in\mathcal U,
        \quad
        t\in\mathcal T,
    \end{equation}
    where $\eta_i^{\mathrm{com}}$ is obtained from either Eq. \eqref{gain-FSO} or Eq. \eqref{gain-fiber}, depending on the underlying optical link. Accordingly, $\tau R_{i,t}$ is the expected number of Bell pairs successfully shared between the QBS and user $i$ during slot $t$.
    
\subsection{Adaptive Multi-GHZ-Probe Preparation With Non-Ideal Fusion}    \label{subsec:ghz_fusion}
    For sensing, elementary Bell pairs are fused into multipartite GHZ probes using the construction in \cite{GHZ-fusion}. The QBS can partition sensing resources among several independent probes per slot, balancing the coherent sensing gain of larger probes against their vulnerability to preparation failures, transmission loss, and decoherence. Under noise, smaller GHZ sub-ensembles can yield higher QFI than a single GHZ state with the same sensing-qubit budget, with the preferred partition depending on the noise conditions \cite{GHZ-partitioning}. Multiple probes also allow successfully prepared and delivered probes to contribute sensing information despite failures of other probes.
    
    Let $\mathcal{G} = \{1,\ldots,G^{\max}\}$ denote the set of candidate GHZ-probe indices, where $G^{\max}$ is a fixed upper bound on the number of probes that can be prepared within one time slot. For every candidate probe $g\in\mathcal{G}$, sensing node $s\in\mathcal{S}$, and slot $t\in\mathcal{T}$, let $n_{s,g,t}\in\mathbb{Z}_{\geq0}$  denote the number of qubits belonging to probe $g$ distributed to sensing node $s$. The probe-level sensing-allocation vector is defined as $\mathbf{n}_{g,t}  =\{n_{s,g,t}\}_{s\in\mathcal{S}}$.     
    Instead of fixing the number of GHZ probes a priori, the number of active GHZ probes adapts to network conditions, available resources, and remaining sensing requirements. It need not be an independent optimization variable, since it is determined by the probe-level qubit allocations $\mathbf{n}_{g,t}$. Specifically, let $m_{g,t}=\sum_{s\in\mathcal{S}}n_{s,g,t}$ denote the number of distributed sensing qubits assigned to candidate probe $g$ in slot $t$. The active probe set is $\mathcal{A}_t=\{g\in\mathcal{G}:m_{g,t}>0\}$, and the number of active probes is $G_t=|\mathcal{A}_t|$.

    Cardinality $G^{\max}$ specifies the maximum number of independently addressable GHZ-probe instances supported by the QBS and, therefore, is a hardware-level bound, rather than a probe-size-dependent resource budget. The number of probes that can actually be activated in a slot is additionally restricted by their sizes through the shared Bell-pair budget and the memory constraints. Specifically, activating larger probes consumes more elementary Bell pairs and, therefore, may reduce the feasible number of simultaneous active probes.
           
    Each active probe retains one idler/reference qubit at the QBS. Therefore, the total number of qubits in active probe  $g\in\mathcal{A}_t$ is $N_{g,t}=1+m_{g,t}$. 
    Following the Bell-pair construction previously described in \cite{GHZ-fusion}, an $N_{g,t}$-partite GHZ state can be assembled from $N_{g,t}-1=m_{g,t}$ elementary Bell pairs and requires $m_{g,t}-1$ fusion operations.  
    To capture non-deterministic fusion in the considered hardware model, each required fusion operation is assumed to independently succeed  with probability $p_{\mathrm{fus}}\in(0,1]$\footnote{Conditional on successful fusion, the resulting GHZ state is assumed ideal; thus, fusion imperfections are represented through the probabilistic preparation-success model in Eq. \eqref{eq:preparation_success}, while additional  gate-induced state infidelity is not considered.}. Accordingly, the preparation-success probability of active probe $g$ is computed as follows:    \begin{equation}
        p_{g,t}^{\mathrm{prep}} = \left(p_{\mathrm{fus}}\right)^{m_{g,t}-1}, \qquad g\in\mathcal A_t.
    \label{eq:preparation_success}
    \end{equation}
    
    An unsuccessful preparation attempt contributes no sensing information; however, the Bell-pair and memory resources reserved for the attempt remain charged to the corresponding slot-level resource budgets. A successfully prepared probe is subsequently affected by transmission loss and memory decoherence, as described in the sensing model.
        
\subsection{Sensing Model}    \label{subsec:sensing_model}
    The sensing subsystem estimates a global scalar parameter associated with a spatially distributed physical field. Parameter $\theta$ is assumed to remain constant over the complete optimization horizon. The local phase shift at sensing node $s$ is $\phi_s(\theta)=a_s\theta$,  where $\theta$ is the unknown parameter, while $a_s\geq0$ is a known location-dependent sensing coefficient.
    
    For each active probe $g\in\mathcal{A}_t$, the QBS attempts to prepare the $N_{g,t}$-qubit GHZ state as follows:
    \begin{equation}
        \ket{\mathrm{GHZ}_{N_{g,t}}}
        =
        \frac{1}{\sqrt{2}}
        \left(
        \ket{0}^{\otimes N_{g,t}}
        +
        \ket{1}^{\otimes N_{g,t}}
        \right). 
        \label{eq:ghz_probe_state}
    \end{equation}
    After successful preparation and delivery, probe $g$ accumulates the phase
    \begin{equation}
        \phi_{g,t}(\theta)
        =
        \theta
        \sum_{s\in\mathcal{S}}
        a_s n_{s,g,t}.
        \label{eq:probe_phase}
    \end{equation}
    Each participating sensing node performs the prescribed local measurement and returns its outcome to the QBS. On receiving the remote outcomes, the QBS measures the stored idler/reference qubit and reconstructs the global parity statistic. We characterize the achievable sensing precision through the QFI, which provides the measurement-independent quantum limit on parameter-estimation performance.    
    Let $\Delta_{g,t}  =  ({ 2\max_{s:n_{s,g,t}>0}d_s})/{c}$ denote the round-trip readout delay, where $c$ is the propagation speed.
    
    \begin{proposition} \label{prop:effective_probe_qfi} 
    Assume that the QBS can determine whether each GHZ-probe preparation and qubit-delivery attempt succeeds or fails. Furthermore, assume that these success or failure events are independent of the unknown parameter $\theta$, and that an  unsuccessfully prepared or delivered probe contributes no sensing information. Let $T_2>0$ denote the idler-memory coherence time under pure dephasing, with other sensing-qubit and measurement imperfections neglected. Then, the effective QFI associated with active probe $g\in\mathcal A_t$ is computed as follows:
    \begin{equation}
    q_{g,t}
    =
    p_{g,t}^{\mathrm{prep}}
    \exp\left(-\frac{2\Delta_{g,t}}{T_2}\right)
    \left(
    \sum_{s\in\mathcal S}
    a_s n_{s,g,t}
    \right)^2
    \prod_{s\in\mathcal S}
    \left(\eta_s^{\mathrm{sen}}\right)^{n_{s,g,t}}.
    \label{eq:effective_probe_qfi}
    \end{equation}
    \end{proposition}

    \begin{IEEEproof}
    Conditional on successful preparation and delivery, and before accounting for idler-memory decoherence, the accumulated phase in Eq. \eqref{eq:probe_phase} yields the GHZ-state QFI as follows:
    \begin{equation}
    F_{g,t}^{\mathrm{succ}}
    =
    \left(
    \sum_{s\in\mathcal S}
    a_s n_{s,g,t}
    \right)^2.
    \label{eq:conditional_probe_qfi}
    \end{equation}
    Idler decoherence over the readout delay $\Delta_{g,t}$ reduces the     off-diagonal coherence by     $\exp(-\Delta_{g,t}/T_2)$ and, consequently, reduces the QFI by     $\exp(-2\Delta_{g,t}/T_2)$.    
    Moreover, since the distributed qubits     are transmitted through independent pure-loss channels, their joint     survival probability is calculated as follows:
    \begin{equation}
    p_{g,t}^{\mathrm{surv}}
    =
    \prod_{s\in\mathcal S}
    \left(\eta_s^{\mathrm{sen}}\right)^{n_{s,g,t}}.
    \label{eq:probe_survival_probability}
    \end{equation}
    Since preparation and delivery outcomes are known to the QBS and their probabilities are independent of $\theta$, the unsuccessful outcomes contain no information about $\theta$. Therefore, only successfully prepared and delivered probes contribute to the sensing information, and their conditional QFI is weighted by the corresponding success probability. Accordingly,
    \begin{equation}\label{eq:effective_qfi_decomposition}
    q_{g,t}
    =
    p_{g,t}^{\mathrm{prep}}
    p_{g,t}^{\mathrm{surv}}
    \exp\left(-\frac{2\Delta_{g,t}}{T_2}\right)
    F_{g,t}^{\mathrm{succ}}.
    \end{equation}
    Substituting Eq.  \eqref{eq:conditional_probe_qfi} and Eq.     \eqref{eq:probe_survival_probability} in Eq.  \eqref{eq:effective_qfi_decomposition} yields  Eq.    \eqref{eq:effective_probe_qfi}.
    \end{IEEEproof}
    
    Here, $q_{g,t}$ is the effective QFI per preparation attempt, averaged over preparation and delivery outcomes, instead of the information realized in each attempt. Since active probes are independently prepared and measured, their QFI contributions are additive. Therefore, the total effective QFI obtained in slot $t$ is computed as follows:
    \begin{equation}
    Q_{\theta,t}^{\mathrm{eff}}
    =
    \sum_{g\in\mathcal A_t}q_{g,t}.
    \label{eq:total_effective_qfi}
    \end{equation}       
               
    As $\theta$ remains constant over the finite horizon, independent measurements collected in different slots correspond to repeated observations of the same parameter, and, accordingly, their QFI contributions are  additive. Quantity $Q_{\theta,t}^{\mathrm{eff}}$ in \eqref{eq:total_effective_qfi} is the \emph{expected} QFI obtained in slot~$t$, where the expectation is taken over the random preparation and delivery outcomes of the active probes. We define $\mathcal{J}_t$ as the expected QFI accumulated before the beginning of slot~$t$. The expected accumulated QFI therefore evolves as follows:
    \begin{equation}
      \mathcal{J}_{t+1}=\mathcal{J}_t+Q_{\theta,t}^{\mathrm{eff}},
      \qquad t\in\mathcal{T},\qquad \mathcal{J}_1=\mathcal{J}_0,
      \label{eq:J_update}
    \end{equation}
    where $\mathcal{J}_0$ denotes initial Fisher information from independent observations, or zero if none are available. Accordingly, the quantum Cram\'er--Rao bound (QCRB) of an unbiased estimator of $\theta$, evaluated at the expected information, is the following equation \cite{variational}:
    \begin{equation}
      \mathrm{QCRB}_t=\frac{1}{\mathcal{J}_t}.
      \label{eq:QCRB}
    \end{equation}
    This quantity is a design-time measure of sensing accuracy. Because the preparation and delivery outcomes are random, the realized accumulated information, denoted by $\mathcal{J}_t^{\mathrm{real}}$, is itself random, and $1/\mathcal{J}_t$ is not a bound on the error of an individual realization. The relation between $\mathcal{J}_{T+1}$ and  $\mathcal{J}_{T+1}^{\mathrm{real}}$ is evaluated numerically in Section \ref{simulation}. 
\section{Dynamic Q-ISAC Resource Allocation Problem}   \label{problem statement}
    In this section, we formulate the finite-horizon joint communication--sensing resource-allocation problem that jointly determines communication Bell-pair allocations and adaptive GHZ-probe configurations in each slot.
\subsubsection{\textbf{Optimization Variables}}
    Optimization variables over the horizon are
    \begin{itemize}[leftmargin=*]
        \item $\mathbf{r}_t=\{r_{i,t}\geq0\}_{i\in\mathcal{U}}$, representing     the continuous communication entanglement-rate allocation; and
        \item $\mathbf{n}_t     =    \{n_{s,g,t}\in\mathbb{Z}_{\geq0}\}_{    s\in\mathcal{S},\,g\in\mathcal{G}}$, denoting the number     of distributed sensing qubits of probe $g$ assigned to
    sensing node $s$ in slot $t$.
    \end{itemize}
    Probe sizes $m_{g,t}$, active-probe set $\mathcal A_t$, number of active probes $G_t$, and total effective QFI $Q_{\theta,t}^{\mathrm{eff}}$ are determined from $\mathbf n_t$ and the corresponding system parameters. 

\subsubsection{\textbf{Constraints}}
    The dynamic Q-ISAC resource-allocation problem is subject to the following constraints:

    \begin{enumerate}[label=(\alph*),leftmargin=*]
    
    \item \textit{Per-Slot Bell-Pair Generation Budget:}
    Since the communication and sensing subsystems share the same entanglement-generation infrastructure, the per-slot Bell-pair consumption must satisfy the following:
    \begin{equation}
        \mathrm{C1}:~~
        \tau\sum_{i\in\mathcal U} r_{i,t}
        +
        \sum_{g\in\mathcal G} m_{g,t}
        \leq N^{\max},
        \qquad
        \forall t\in\mathcal T.
        \label{eq:entanglement_budget}
    \end{equation}
    
    \item \textit{Time-Averaged Communication QoS:}
    Each communication user $i\in\mathcal U$ must achieve a minimum time-averaged delivered entanglement rate $R_i^{\min}$ over the optimization horizon:
    \begin{equation}
        \mathrm{C2}:~~
        \frac{1}{T}\sum_{t\in\mathcal T} R_{i,t}
        \geq
        R_i^{\min},
        \qquad
        \forall i\in\mathcal U.
        \label{eq:average_qos}
    \end{equation}


   \item   \textit{Terminal Sensing Requirement:} Let $\epsilon^{\max}>0$ denote the maximum allowable terminal QCRB, evaluated at the expected accumulated information. The sensing requirement is $\mathrm{QCRB}_{T+1}\le\epsilon^{\max}$. Since $\mathcal{J}_{T+1}>0$ and using \eqref{eq:QCRB}, this condition is equivalently written as follows:
    \begin{equation}
      \mathrm{C3:}\quad \mathcal{J}_{T+1}\ge\frac{1}{\epsilon^{\max}}.
      \label{eq:terminal_sensing}
    \end{equation}

    \item \textit{Per-Slot Sensing Reliability:} 
    Along with the terminal sensing requirement, the sensing subsystem must provide a minimum amount of sensing information with sufficiently high probability in each slot. Let $Z_{g,t}\in\{0,1\}$ be a Bernoulli random variable indicating the successful preparation and delivery of probe $g$, where $Z_{g,t}\sim  \mathrm{Bernoulli}\!\left(  p_{g,t}^{\mathrm{prep}}p_{g,t}^{\mathrm{surv}} \right)$ and variables $\{Z_{g,t}\}_{g\in\mathcal A_t}$ are assumed mutually independent. Accordingly, the realized QFI in slot $t$ is computed as follows:
    \begin{equation}
    Q_{\theta,t}^{\mathrm{real}}
    =
    \sum_{g\in\mathcal A_t}
    Z_{g,t}
    \exp\!\left(-\frac{2\Delta_{g,t}}{T_2}\right)
    \left(
    \sum_{s\in\mathcal S}a_s n_{s,g,t}
    \right)^2 .
    \label{eq:realized_qfi}
    \end{equation}
    The per-slot sensing-reliability requirement is then
    \begin{equation}
        \mathrm{C4}:~~
        \Pr\!\left(
        Q_{\theta,t}^{\mathrm{real}}
        \geq Q^{\min}_{\mathrm{slot}}
        \right)
        \geq 1-\delta_s,
        \qquad \forall t\in\mathcal T,
        \label{eq:sensing_reliability}
    \end{equation}
    where $\delta_s\in[0,1)$ is the allowable sensing-outage probability and the per-slot sensing threshold is selected as $Q_{\mathrm{slot}}^{\min}    =    {\mathcal J^{\mathrm{target}}}/{T}$.  $\mathrm{C3}$ constrains the \emph{expected} terminal information, whereas $\mathrm{C4}$ controls the probability of insufficient \emph{realized} information in each slot. 

    \item \textit{Communication-User Processing Capacity:} 
    Let $C_i^{\mathrm{com}}$ denote the per-slot receiver processing capacity. We constrain the expected number of delivered Bell-pair halves as follows:
    \begin{equation}
        \mathrm{C5}:~~
        \tau r_{i,t}\eta_i^{\mathrm{com}}
        \leq
        C_i^{\mathrm{com}},
        \qquad
        \forall i\in\mathcal U,\quad
        \forall t\in\mathcal T.
        \label{eq:communication_processing_capacity}
    \end{equation}
    
    \item \textit{Sensing-Node Processing Capacity:}
    Each sensing node has a finite capability to receive, manipulate, and measure sensing qubits within one slot. Let $C_s^{\mathrm{sen}}$ denote the maximum number of sensing qubits that sensing node $s\in\mathcal S$ can process during a slot. Accordingly,
    \begin{equation}
        \mathrm{C6}:~~
        \sum_{g\in\mathcal G} n_{s,g,t}
        \leq
        C_s^{\mathrm{sen}},
        \qquad
        \forall s\in\mathcal S,\quad
        \forall t\in\mathcal T.
        \label{eq:sensing_processing_capacity}
    \end{equation}
    
    \item \textit{QBS Quantum-Memory Capacity:}
    For every communication Bell pair allocated in a slot, one qubit is retained at the QBS. 
    In addition, each active GHZ probe requires one idler qubit to remain stored at the QBS during the sensing readout procedure. Accordingly, the transient QBS memory occupancy must satisfy the following:
    \begin{equation}
        \mathrm{C7}:~~
        \tau\sum_{i\in\mathcal U} r_{i,t}
        +
        G_t
        \leq
        M_0,
        \qquad
        \forall t\in\mathcal T,
        \label{eq:qbs_memory_dynamic}
    \end{equation}
    where $M_0$ denotes the number of quantum-memory qubits available at the QBS. The QBS memory is modeled as a per-slot working buffer.
    
    \item \textit{Fusion-Processing Capacity:}
    Since the QBS can execute only a finite number of fusion operations within each slot, let $F^{\max}$ denote its per-slot fusion-processing capacity. The total number of required fusion operations must satisfy the following constraint:
    \begin{equation}
        \mathrm{C8}:~~
        \sum_{g\in\mathcal A_t}
        \left(m_{g,t}-1\right)
        \leq
        F^{\max},
        \qquad
        \forall t\in\mathcal T.
        \label{eq:fusion_capacity}
    \end{equation}
    \end{enumerate}
\subsubsection{\textbf{Objective Function}}
    The QBS allocates its finite Bell-pair generation capacity to maximize the cumulative $\rho$-fair utility of the communication users' slot-level delivered entanglement rates, while satisfying the communication QoS and sensing requirements. Otherwise, users with more favorable communication channels may receive an unfairly large share of the residual Bell-pair capacity after the minimum QoS requirements are satisfied. To mitigate excessive resource concentration and balance aggregate entanglement throughput with service fairness, we adopt the per-user utility as follows:
    \begin{equation}
        U_{\rho}(R_{i,t}) =  \dfrac{R_{i,t}^{1-\rho}}{1-\rho}, \qquad  \rho \in \left(0,1\right), \quad \forall i \in\mathcal{U}.
        \label{eq:rho_fair_utility}
    \end{equation}
    
    Based on the definition of decision variables, constraints, and the objective function, the dynamic Q-ISAC resource-allocation problem is formulated as follows:
    \begin{equation}
    \label{eq:dynamic_qisac_problem}
    \begin{aligned}
        \max_{\{\mathbf r_t,\mathbf n_t\}_{t\in\mathcal T}}
        \dfrac{1}{T}\sum_{t\in\mathcal T}\sum_{i\in\mathcal U}U_{\rho}(R_{i,t}) \qquad     
        \mathrm{s.t.} \quad   \mathrm{C1}-\mathrm{C8}.
    \end{aligned}
    \end{equation}

\section{Proposed Hierarchical Constrained PPO Solution}\label{algorithm}

    The dynamic Q-ISAC resource-allocation problem in Eq. \eqref{eq:dynamic_qisac_problem} is a finite-horizon mixed-integer nonlinear problem with per-slot sensing-reliability chance constraints. Its difficulty arises from the combinatorial sensing allocation, the nonlinear dependence of QFI and reliability on GHZ-probe configurations, and the coupling of decisions across the horizon.

    Although offline optimization provides benchmarks for tractable instances, its computational burden can grow rapidly with network size and horizon length, and optimization must be repeated when the system state or operating requirements change. This motivates HC-PPO, which shifts the main computational burden to offline training and enables efficient online sensing-allocation decisions. 
    To address Problem \eqref{eq:dynamic_qisac_problem}, we decompose each slot into a discrete sensing-allocation stage and a continuous communication-allocation stage. The sensing decision determines the resources remaining for communication, after which the communication subproblem reduces to a concave $\rho$-fair rate-allocation problem with linear constraints, solved analytically through water filling. The sensing stage is formulated as a finite-horizon constrained Markov decision process (CMDP) and addressed by a constrained PPO method \cite{CPPO}. Constraints $\mathrm{C1}$ and $\mathrm{C5}$--$\mathrm{C8}$ are enforced through action masking and the communication allocator; communication reachability checks address $\mathrm{C2}$, while a joint sensing shield attempts to repair allocations violating $\mathrm{C3}$ or $\mathrm{C4}$. Projected Lagrange multipliers penalize the corresponding horizon-and slot-level residuals during training.

\subsection{Learning-Based Sensing Allocation}\label{subsec:sensing_solution}
    The sensing-allocation problem induced by the analytical communication allocator is written as follows:
    \begin{equation}
    \label{eq:sensing_problem}
    \begin{aligned}
    \max_{\{\mathbf n_t\}_{t\in\mathcal T}}
    &\sum_{t\in\mathcal T}\sum_{i\in\mathcal U}
    U_{\rho}\!\left(R_{i,t}^{\mathrm{wf}}\right)\quad 
    \mathrm{s.t.}\quad
    &\mathrm{C1}-\mathrm{C4},~~\mathrm{C6}-\mathrm{C8},
    \end{aligned}
    \end{equation}
    where $R_{i,t}^{\mathrm{wf}} =\eta_i^{\mathrm{com}}r_{i,t}^{\mathrm{wf}}$ is the delivered rate returned by the water-filling allocator for a fixed $\mathbf n_t$. The communication-related portions of shared constraints $\mathrm{C1}$ and $\mathrm{C7}$, together with $\mathrm{C5}$, are enforced by the analytical allocator. The remaining constraints are handled through action masking, reachability tests, joint sensing shield, and projected Lagrange multipliers, as detailed below.

\begin{enumerate}[leftmargin=*]
    \item \textit{CMDP Formulation:}
        At slot $t$, define user $i$'s remaining communication requirement as follows:
        \begin{equation}
        D_{i,t}
        =
        \left[
        TR_i^{\min}
        -
        \sum_{\ell=1}^{t-1}R_{i,\ell}
        \right]^+,
        \qquad i\in\mathcal U,
        \label{eq:remaining_communication_requirement}
        \end{equation}
        and let $k_t=T-t+1$. The agent state is then
        \begin{equation}
        \mathbf s_t=
        \left[
        \boldsymbol{\eta}^{\mathrm{com}},
        \boldsymbol{\eta}^{\mathrm{sen}},
        \mathbf a,
        \widetilde{\mathbf d},
        \widetilde{\mathcal J}_t,
        \widetilde{\mathbf D}_t,
        \frac{k_t}{T}
        \right],
        \label{eq:cmdp_state}
        \end{equation}
        where
        \begin{equation}
        \widetilde{\mathcal J}_t=\epsilon^{\max}\mathcal J_t,
        \qquad
        \widetilde D_{i,t}
        =
        \frac{D_{i,t}}{TR_i^{\min}},
        \quad
        \widetilde{\mathbf D}_t
        =
        [\widetilde D_{i,t}]_{i\in\mathcal U}. 
        \label{eq:normalized_state_variables}
        \end{equation}
        Also, in Eq. \eqref{eq:cmdp_state}, $\mathbf a=[a_s]_{s\in\mathcal S}$ collects sensing coefficients and $\widetilde{\mathbf d}=[\widetilde d_s]_{s\in\mathcal S}$ denotes the normalized sensing-node distance vector, with $\widetilde d_s=d_s/d^{\max}$.
        Furthermore, the normalized base reward is computed as follows:
        \begin{equation}
        r_t^{\mathrm{base}}
        =
        \frac{\sum_{i\in\mathcal U}U_\rho(R_{i,t})}{f^{\max}},
        \label{eq:sensing_base_reward}
        \end{equation}
        where $f^{\max}>0$ is a normalization factor. The signed residuals of $\mathrm{C2}$ and $\mathrm{C3}$ are then
        \begin{align}
        c_i^{\mathrm{com}}
        &=
        \frac{
        R_i^{\min}
        -
        \frac{1}{T}\sum_{t=1}^{T}R_{i,t}
        }{
        R_i^{\min}
        },
        \label{eq:communication_terminal_residual}\\
        c^{\mathrm{sen}}
        &=
        \frac{
        \mathcal J^{\mathrm{target}}-\mathcal J_{T+1}
        }{
        \mathcal J^{\mathrm{target}}
        }
        =
        1-\epsilon^{\max}\mathcal J_{T+1},
        \label{eq:sensing_terminal_residual}
        \end{align}
        where $\mathcal{J}^{\mathrm{target}}=1/\epsilon^{\max}$ and $\mathcal{J}_{T+1}$ is the expected terminal information.  Since $\mathrm{C4}$ is imposed per slot, its residual is calculated as follows:
        \begin{equation}
        c_t^{\mathrm{rel}}
        =
        \frac{
        (1-\delta_s)
        -
        \Pr\!\left(
        Q_{\theta,t}^{\mathrm{real}}
        \geq Q_{\mathrm{slot}}^{\min}
        \right)
        }{
        1-\delta_s
        }.
        \label{eq:reliability_residual}
        \end{equation}
        Positive residuals indicate violations. The augmented PPO reward is thus
        \begin{equation}
        \widehat r_t
        =
        r_t^{\mathrm{base}}
        -
        \mathbf 1_{\{t=T\}}
        \left(
        \sum_{i\in\mathcal U}\xi_i c_i^{\mathrm{com}}
        +
        \mu c^{\mathrm{sen}}
        \right)
        -
        \lambda c_t^{\mathrm{rel}},
        \label{eq:sensing_lagrangian_reward}
        \end{equation}
        where $\xi_i,\mu,\lambda\geq0$ are the multipliers associated with $\mathrm{C2}$, $\mathrm{C3}$, and $\mathrm{C4}$, respectively.
    \item \textit{Masked autoregressive sensing actor:}
        To handle the combinatorial sensing-action space, we adopt a sequential autoregressive construction. In so doing, we follow the general idea of sequential neural combinatorial optimization \cite{sequential-action}, along with state-dependent invalid-action masking \cite{masking}.
        To avoid representing the combinatorial sensing-allocation matrix as a single categorical action, the actor sequentially constructs $\mathbf n_t^\pi$  as shown below:
        \begin{equation}
        \boldsymbol{\alpha}_t= (a_t^{(1)},\ldots,a_t^{(L_t)}), \qquad \mathbf n_t^\pi=\mathcal D(\boldsymbol{\alpha}_t),
        \label{eq:autoregressive_action_sequence}
        \end{equation}
        where $L_t\leq L$ is the sequence length and $a_t^{(\ell)} \in  \{\mathtt{STOP}\}  \cup  \{(g,s,q):  g\in\mathcal G,\, s\in\mathcal S,\,  q\in\{1,\ldots,q^{\max}\}\}$, in which  $q^{\max}$ denotes the maximum number of sensing qubits added by a single elementary action.
        Action $(g,s,q)$ adds $q$ sensing qubits of probe $g$ to node $s$, $\mathtt{STOP}$ terminates the construction, and $\mathcal D(\cdot)$ maps the resulting action sequence to the allocation matrix.          
        The policy factorizes as follows:
        \begin{equation}
        \pi_{\boldsymbol{\theta}}
        (\boldsymbol{\alpha}_t|\mathbf s_t)
        =
        \prod_{\ell=1}^{L_t}
        \pi_{\boldsymbol{\theta}}
        \!\left(
        a_t^{(\ell)}
        \mid
        \mathbf s_t,\mathbf n_t^{(\ell-1)}
        \right).
        \label{eq:autoregressive_policy}
        \end{equation}
        
        A state-dependent mask removes candidate increments violating the instantaneous sensing-side constraints or render the remaining communication QoS requirements unreachable. For a candidate partial allocation $\widehat{\mathbf n}$, with $\widehat m_{g,t}=\sum_s\widehat n_{s,g,t}$ and $\widehat G_t$ active probes, the mask requires
        \begin{equation*}
        \begin{aligned}
        &\sum_g\widehat n_{s,g,t}\leq C_s^{\mathrm{sen}},\quad \forall s,
        &&
        \sum_{g,s}\widehat n_{s,g,t}\leq N^{\max},\\
        &\widehat G_t\leq\min\{G^{\max},M_0\},
        &&
        \sum_g\widehat m_{g,t}-\widehat G_t\leq F^{\max}.
        \end{aligned}
        \end{equation*}
        
        The communication capacity remaining after the candidate sensing action is computed as follows:
        \begin{equation}
        \widehat B_t
        =
        \min\left\{
        N^{\max}-\sum_{g,s}\widehat n_{s,g,t},
        \,
        M_0-\widehat G_t
        \right\}.
        \label{eq:candidate_communication_budget}
        \end{equation}
        
        \begin{lemma}
        \label{lem:communication_reachability}
        Under quasi-static communication channels, a feasible continuation satisfying $\mathrm{C2}$ can exist only if
        \begin{equation}
        D_{i,t}
        \leq
        k_t\frac{C_i^{\mathrm{com}}}{\tau},
        \qquad \forall i\in\mathcal U,
        \label{eq:individual_communication_reachability}
        \end{equation}
        and
        \begin{equation}
        \sum_{i\in\mathcal U}
        \frac{D_{i,t}}{\eta_i^{\mathrm{com}}}
        \leq
        \frac{\widehat B_t}{\tau}
        +
        (k_t-1)
        \frac{\min\{N^{\max},M_0\}}{\tau}.
        \label{eq:aggregate_communication_reachability}
        \end{equation}
        \end{lemma}
        
        \begin{IEEEproof}
        Constraint $\mathrm{C5}$ limits the remaining service of user $i$ to $k_tC_i^{\mathrm{com}}/\tau$, yielding Eq. \eqref{eq:individual_communication_reachability}. Moreover, $\mathrm{C1}$ and $\mathrm{C7}$ limit the current communication generation rate to $\widehat B_t/\tau$. Optimistically assuming that no sensing resources are used thereafter gives at most $\min\{N^{\max},M_0\}/\tau$ per future slot. Since delivering $D_{i,t}$ requires $D_{i,t}/\eta_i^{\mathrm{com}}$ generated Bell pairs, Eq. \eqref{eq:aggregate_communication_reachability} follows.
        \end{IEEEproof}
         Candidates violating either condition are masked. These conditions are necessary, but not sufficient for terminal satisfaction of $\mathrm{C2}$. Likewise, the sensing-reachability bound in Lemma 2 is necessary but not sufficient for $\mathrm{C3}$, and the shield is a heuristic repair: it may return an allocation that still violates $\mathrm{C3}$ or $\mathrm{C4}$, in which case the transition is marked shield-infeasible. The masks and the shield therefore do not guarantee joint feasibility; feasibility is assessed empirically on held-out instances in Section \ref{simulation}.  
    \item \textit{Joint sensing-reachability and reliability shield:}
        Inspired by safe reinforcement learning via shielding \cite{shield}, we introduce a deterministic problem-specific shield that replaces actor-generated sensing allocations when  sensing reachability or reliability conditions are violated.
        For an allocation $\mathbf n_t$, let us define $P_t^{\mathrm{rel}}(\mathbf n_t)=\Pr\!\left( Q_{\theta,t}^{\mathrm{real}} \geq Q_{\mathrm{slot}}^{\min} \right)$. Termination normally requires a complete probe partition satisfying both sensing conditions; if no feasible increment remains, the resulting allocation is passed to the fallback shield.        
        An optimistic one-slot QFI upper bound is obtained from the following:
        \begin{equation}
        \begin{aligned}
           & N_{\mathrm{sen}}^{\max}
        =
        \min\left\{
        N^{\max},
        \sum_sC_s^{\mathrm{sen}},
        F^{\max}+G^{\max}
        \right\},\\
        &\overline Q
        =
        \left(
        N_{\mathrm{sen}}^{\max}\max_s|a_s|
        \right)^2.
        \end{aligned}
        \label{eq:qfi_upper_bound}
        \end{equation}
        
        \begin{lemma}
        \label{lem:sensing_reachability}
        A necessary condition for $\mathcal J_{T+1}\geq\mathcal J^{\mathrm{target}}$ to remain reachable after selecting $\mathbf n_t$ is
        \begin{equation}
        Q_{\theta,t}^{\mathrm{eff}}(\mathbf n_t)
        \geq
        Q_t^{\mathrm{req}},
        \qquad
        Q_t^{\mathrm{req}}
        =
        \left[
        \mathcal J^{\mathrm{target}}
        -
        \mathcal J_t
        -
        (k_t-1)\overline Q
        \right]^+.
        \label{eq:required_current_qfi}
        \end{equation}
        \end{lemma}
        
        \begin{IEEEproof}
        Even if every future slot achieves $\overline Q$,  $\mathcal J_{T+1} \leq \mathcal J_t + Q_{\theta,t}^{\mathrm{eff}}(\mathbf n_t) + (k_t-1)\overline Q$, which directly yields Eq. \eqref{eq:required_current_qfi}.
        \end{IEEEproof}
        
        If actor allocation $\mathbf n_t^\pi$ violates either $Q_{\theta,t}^{\mathrm{eff}}(\mathbf n_t^\pi)\geq Q_t^{\mathrm{req}}$ or $P_t^{\mathrm{rel}}(\mathbf n_t^\pi)\geq1-\delta_s$, the shield constructs a replacement allocation from $\mathbf n=\mathbf0$.
        Let us define
        \begin{align}
        \gamma_Q(\mathbf n)
        &=
        \begin{cases}
        1, & Q_t^{\mathrm{req}}=0,\\
        \min\!\left\{
        Q_{\theta,t}^{\mathrm{eff}}(\mathbf n)/Q_t^{\mathrm{req}},1
        \right\},
        & Q_t^{\mathrm{req}}>0,
        \end{cases}
        \label{eq:qfi_progress}\\
        \gamma_R(\mathbf n)
        &=
        \min\!\left\{
        P_t^{\mathrm{rel}}(\mathbf n)/(1-\delta_s),1
        \right\}.
        \label{eq:reliability_progress}
        \end{align}
        
        For each feasible candidate $a\in\mathcal F_t(\mathbf n)$ with $\mathbf n'=\mathbf n\oplus a$, the shield lexicographically maximizes as follows:
        \begin{equation}
        \mathbf z(a)
        =
        \left[
        \Delta\min\{\gamma_Q,\gamma_R\},
        ~
        \Delta\gamma_Q+\Delta\gamma_R,
        ~
        \Delta Q_{\theta,t}^{\mathrm{eff}},
        ~
        -\|\mathbf n'\|_1
        \right]^{\mathsf T},
        \label{eq:joint_shield_score}
        \end{equation}
        where each $\Delta$ denotes the corresponding change from $\mathbf n$ to $\mathbf n'$. Thus,
        \begin{equation}
        a^{\mathrm{sh}}
        =
        \operatorname*{arg\,max}^{\mathrm{lex}}_
        {a\in\mathcal F_t(\mathbf n)}
        \mathbf z(a),
        \qquad
        \mathbf n\leftarrow\mathbf n\oplus a^{\mathrm{sh}}.
        \label{eq:greedy_shield_action}
        \end{equation}
        The construction stops when both sensing conditions are met, no improving feasible action remains, or the construction-depth limit is reached. If either sensing condition remains violated at termination, the transition is marked shield-infeasible.      
    \item \textit{Constrained PPO training:}
        Let $\mathcal I_{\mathrm{ns}}$ denote transitions without a shield override. For such transitions,
        \begin{equation}
        \chi_t(\boldsymbol{\theta})
        =
        \frac{
        \pi_{\boldsymbol{\theta}}
        (\boldsymbol{\alpha}_t|\mathbf s_t)
        }{
        \pi_{\boldsymbol{\theta}_{\mathrm{old}}}
        (\boldsymbol{\alpha}_t|\mathbf s_t)
        },
        \label{eq:ppo_probability_ratio}
        \end{equation}
        and the PPO objective is $\mathcal J_{\mathrm{clip}}         =         \mathbb E_{t\in\mathcal I_{\mathrm{ns}}}         \!\left[         \min         \left\{         \chi_t\widehat A_t,         \operatorname{clip}         (\chi_t,1-\epsilon_{\mathrm{PPO}},1+\epsilon_{\mathrm{PPO}})         \widehat A_t         \right\}         \right]$, in which $\widehat A_t$ is computed using generalized advantage estimation from the augmented rewards in Eq. \eqref{eq:sensing_lagrangian_reward}. Shielded transitions are excluded from the PPO surrogate objective but retained for critic and constraint-residual estimation. Shielded transitions passing the necessary $\mathrm{C3}$ reachability check and satisfying $\mathrm{C4}$ form $\mathcal I_{\mathrm{sh}}$ and provide behavior-cloning targets:
        \begin{equation}
        \mathcal L_{\mathrm{BC}}
        =
        -
        \mathbb E_{t\in\mathcal I_{\mathrm{sh}}}
        \left[
        \log 
        \pi_{\boldsymbol{\theta}}
        (\boldsymbol{\alpha}_t^{\mathrm{sh}}|\mathbf s_t)
        \right].
        \label{eq:shield_behavior_cloning_loss}
        \end{equation}
        
        The critic and total losses are computed as follows:
        \begin{align}
        \mathcal L_V
        &=
        \mathbb E_t
        \left[
        \left(
        V_{\boldsymbol{\psi}}(\mathbf s_t)-\widehat V_t
        \right)^2
        \right],
        \label{eq:value_loss}\\
        \mathcal L
        &=
        -\mathcal J_{\mathrm{clip}}
        +c_V\mathcal L_V
        -c_H\mathcal H
        +\beta_{\mathrm{BC}}\mathcal L_{\mathrm{BC}},
        \label{eq:total_ppo_loss}
        \end{align}
        where $\mathcal H$ is the policy entropy,  $V_{\boldsymbol{\psi}}(\mathbf s_t)$ is the critic parameterized by $\boldsymbol{\psi}$, and $\widehat V_t$ is the return target used for critic training, while $c_V$, $c_H$, and $\beta_{\mathrm{BC}}$ are the critic-loss, entropy, and behavior-cloning coefficients, respectively.
        
        Finally, the projected multiplier updates are computed as follows:
        \begin{align}
        \xi_i
        &\leftarrow
        \Pi_{[0,\Lambda^{\max}]}
        \!\left[
        \xi_i+\alpha_{\mathrm{com}}\overline c_i^{\mathrm{com}}
        \right],
        \qquad \forall i\in\mathcal U,
        \label{eq:communication_dual_update}\\
        \mu
        &\leftarrow
        \Pi_{[0,\Lambda^{\max}]}
        \!\left[
        \mu+\alpha_{\mathrm{sen}}\overline c^{\mathrm{sen}}
        \right],
        \label{eq:sensing_dual_update}\\
        \lambda
        &\leftarrow
        \Pi_{[0,\Lambda^{\max}]}
        \!\left[
        \lambda+\alpha_{\mathrm{rel}}\overline c^{\mathrm{rel}}
        \right].
        \label{eq:reliability_dual_update}
        \end{align}
        Here, $\overline c_i^{\mathrm{com}}$ and $\overline c^{\mathrm{sen}}$ are averaged over completed episodes, whereas $\overline c^{\mathrm{rel}}$ is averaged over sampled slot transitions.

        In Eqs. \eqref{eq:communication_dual_update}--\eqref{eq:reliability_dual_update}, $\Pi_{[0,\Lambda^{\max}]}[\cdot]$ denotes projection onto $[0,\Lambda^{\max}]$, where $\Lambda^{\max}$ is the maximum allowed dual-variable value. Parameters $\alpha_{\mathrm{com}}$,  $\alpha_{\mathrm{sen}}$, and $\alpha_{\mathrm{rel}}$ are the dual learning rates associated with C2, C3, and C4, respectively.
        
\end{enumerate} 
\subsection{QoS-Aware Communication Allocation}\label{subsec:communication_solution}
    After sensing allocation $\mathbf n_t$ is finalized, the residual communication generation-rate capacity is computed as shown below:
    \begin{equation}
    C_t^{\mathrm{com}}(\mathbf n_t) = \frac{1}{\tau} \min \left\{ N^{\max} - \sum_{g\in\mathcal G}\sum_{s\in\mathcal S}n_{s,g,t}, \; M_0-G_t \right\}.
    \label{eq:residual_communication_capacity}
    \end{equation}    
    Given $\mathbf n_t$, the communication rates are obtained by solving the following:
    \begin{equation}
    \label{eq:communication_subproblem}
    \begin{aligned} 
    \underset{\mathbf r_t}{\max} \quad
    & \sum_{i\in\mathcal U} U_\rho\!\left(\eta_i^{\mathrm{com}}r_{i,t}\right) \\
    \operatorname{s.t.}\quad
    &\underline r_{i,t} \leq r_{i,t} \leq u_i = \frac{C_i^{\mathrm{com}}} {\tau\eta_i^{\mathrm{com}}}, \qquad i\in\mathcal U, \\
    &\sum_{i\in\mathcal U}r_{i,t} \leq C_t^{\mathrm{com}}(\mathbf n_t).
    \end{aligned}
    \end{equation}
    Here, the per-user upper bounds enforce  communication-user processing constraint $\mathrm{C5}$, whereas the lower bounds provide QoS-reachability floors. The aggregate rate constraint represents the residual forms of $\mathrm{C1}$ and $\mathrm{C7}$.
    The initial rate floor required by the future per-user processing limit is thus
    \begin{equation}
    \underline r_{i,t}^{(0)} = \frac{ \left[ D_{i,t} - (k_t-1)\dfrac{C_i^{\mathrm{com}}}{\tau} \right]^+ }{ \eta_i^{\mathrm{com}} }.
    \label{eq:initial_communication_rate_floor}
    \end{equation}
    When necessary, these floors are increased according to the urgency index: $\upsilon_{i,t} = {D_{i,t}}/ \left({k_tC_i^{\mathrm{com}}/\tau}\right)$, subject to the available current-slot capacity. If the resulting floors are infeasible, an instantaneously feasible best-effort allocation is applied.
    
    \begin{proposition} \label{prop:waterfilling}
    For $0<\rho<1$, if Problem \eqref{eq:communication_subproblem} is feasible, its optimal solution is
    \begin{equation}
    r_{i,t}^{\star} = \left[ \left(  \frac{(\eta_i^{\mathrm{com}})^{1-\rho}} {\nu_t} \right)^{1/\rho} \right]_{\underline r_{i,t}}^{u_i}, \qquad i\in\mathcal U,
    \label{eq:waterfilling_solution}
    \end{equation} 
    If $C_t^{\mathrm{com}}(\mathbf n_t)\geq\sum_i u_i$, then $r_{i,t}^{\star}=u_i$ for all users. Otherwise, $\nu_t>0$ is selected such that $\sum_i r_{i,t}^{\star}= C_t^{\mathrm{com}}(\mathbf n_t)$.
    \end{proposition}
        
    \begin{IEEEproof}
    The objective is concave for $0<\rho<1$ and the constraints are affine. The KKT stationarity condition for an interior solution is $(\eta_i^{\mathrm{com}})^{1-\rho} r_{i,t}^{-\rho} = \nu_t$. Solving for $r_{i,t}$ and applying the box constraints yields Eq. \eqref{eq:waterfilling_solution}. Water level $\nu_t$ is obtained through one-dimensional bisection.
    \end{IEEEproof}   
    
    The delivered rate is $R_{i,t} = \eta_i^{\mathrm{com}}r_{i,t}^{\star}, \qquad i\in\mathcal U$ and the state variables evolve as follows:
    \begin{align}
    \mathcal J_{t+1} &= \mathcal J_t+ Q_{\theta,t}^{\mathrm{eff}}(\mathbf n_t), \label{eq:updating-J}\\
    D_{i,t+1} &=\left[D_{i,t}-R_{i,t}\right]^+ \label{eq:D_update}.
    \end{align}
    These updates track expected accumulated sensing information and remaining expected communication-rate requirements. Although probe outcomes are heralded at the QBS, the allocation controller uses expected information instead of realized probe outcomes. Thus, $\widetilde{\mathcal J}_t =\epsilon^{\max}\mathcal J_t$ is the normalized expected sensing information. 
\section{Simulation Results} \label{simulation}
    In this section, we report the results of evaluating the proposed HC-PPO framework and investigating the benefits of adaptive GHZ-probe partitioning and joint communication--sensing resource allocation. We first examine the convergence and constraint-satisfaction behavior of HC-PPO and then compare the proposed scheme with separable-probing,  single-GHZ, fixed resource-splitting, and offline optimization benchmarks under different system configurations. Numerical results are reported for both FSO and fiber-optic channels.
\subsection{Simulation Setup}
    We considered a Q-ISAC network with a QBS located at the center of a     $200\times200$ m$^2$ deployment region. Letting $A=200$ m denote the     side length, the Cartesian coordinates of the communication users were     independently drawn from $ \mathrm{Unif}[0.05A,0.95A]$, while those of the     sensing nodes were drawn from $\mathrm{Unif}  [0.25A,0.85A]$. Unless otherwise     stated, the network comprised $U=10$ communication users, $S=5$ sensing     nodes, and at most $G^{\max}=4$ active GHZ probes over a horizon of     $T=8$ slots with normalized slot duration $\tau=1$.     
    The QBS had a per-slot Bell-pair generation budget of     $N^{\max}=1000$, quantum-memory capacity $M_0=1000$, and     fusion-processing capacity $F^{\max}=1000$. The communication-user and     sensing-node processing capacities are $C_i^{\rm com}=200$ and     $C_s^{\rm sen}=75$, respectively. Unless varied in a particular     experiment, we set $R_i^{\min}=45$,     $\mathcal J^{\rm target}=1/\epsilon^{\max}=25$,     $\delta_s=0.1$, $Q_{\rm slot}^{\min}=\mathcal J^{\rm target}/T=3.125$,     $p_{\rm fus}=0.95$, $T_2=10^{-3}$ s, and $\rho=0.5$.  The sensing coefficients were independently generated as     $a_s\sim\mathrm{Unif}  [0.8,1.2]$.
    
    For FSO links, we used an optical wavelength of $1550$ nm, atmospheric     attenuation of $0.1$ dB/km, beam divergence of $8$ mrad, receiver     aperture radius of $0.55$ m, and angular-jitter variance of     $2\times10^{-9}$ rad$^2$. For fiber links, the attenuation coefficient     amounted to $0.2$ dB/km. Channel transmissivities were calculated according to     the models described in Section~\ref{channels}. 
\subsection{Convergence and Constraint Satisfaction of HC-PPO} 
\begin{figure}
        \centering
        \includegraphics[width=0.775\linewidth]{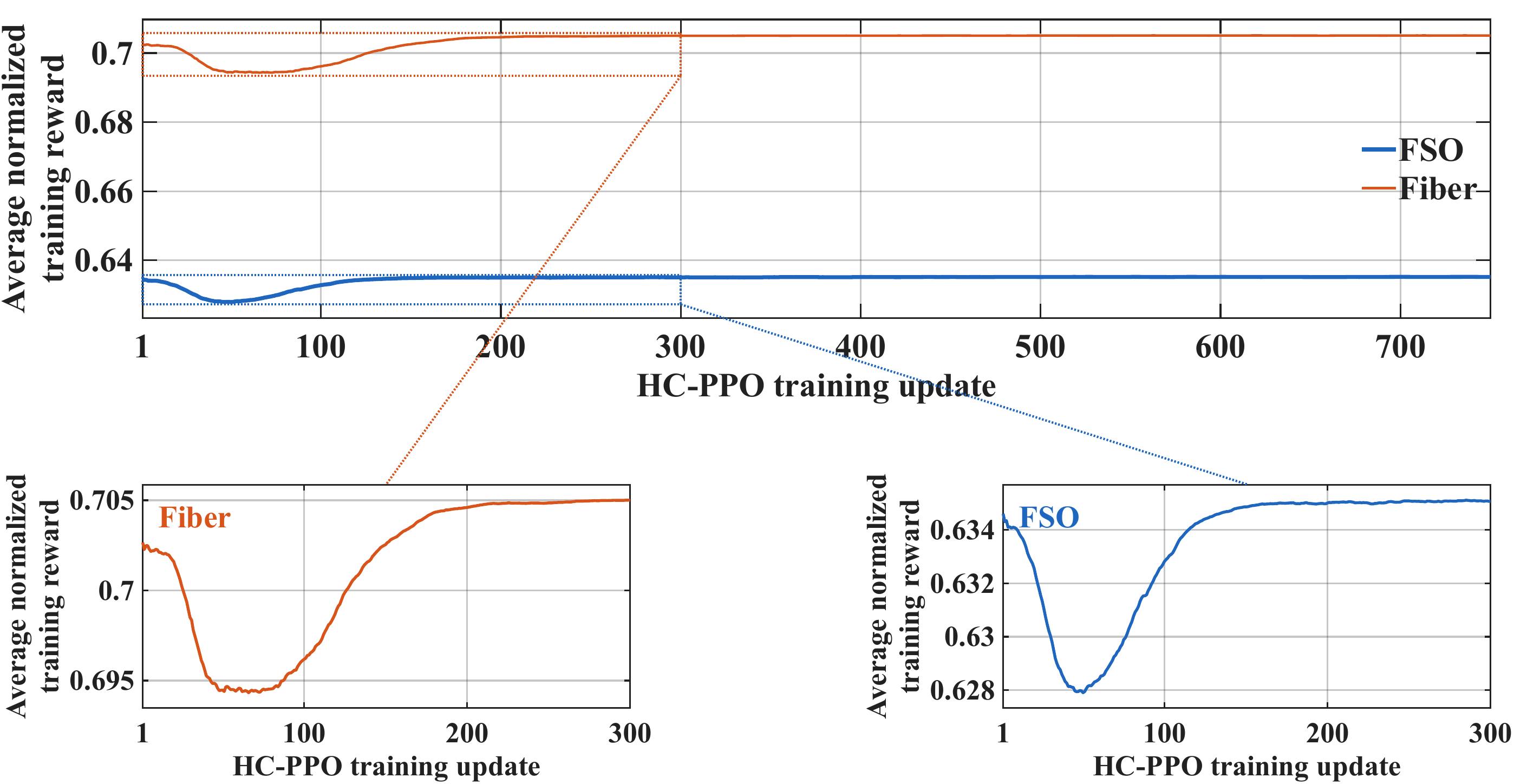}
        \caption{\small Convergence behavior of the proposed HC-PPO algorithm for both fiber and FSO channels. }
        \label{fig:convergence}
    \end{figure}

    Fig. \ref{fig:convergence} illustrates the convergence behavior of HC-PPO over $750$ training updates for FSO and fiber channels. The normalized training reward initially decreased slightly and subsequently stabilized. For FSO, it reached a minimum of approximately $0.628$ before converging around $0.635$, whereas, for fiber, it increased from approximately $0.693$ toward $0.705$. On the independent test set of $750$ instances per training seed, the normalized reward was $0.636$ for FSO and $0.705$ for fiber. Under the considered settings, the higher reward obtained over fiber links was consistent with their higher transmissivity, whereas the FSO links additionally experienced beam-spreading and pointing losses. 
    
\begin{figure}
        \centering
        \includegraphics[width=0.775\linewidth]{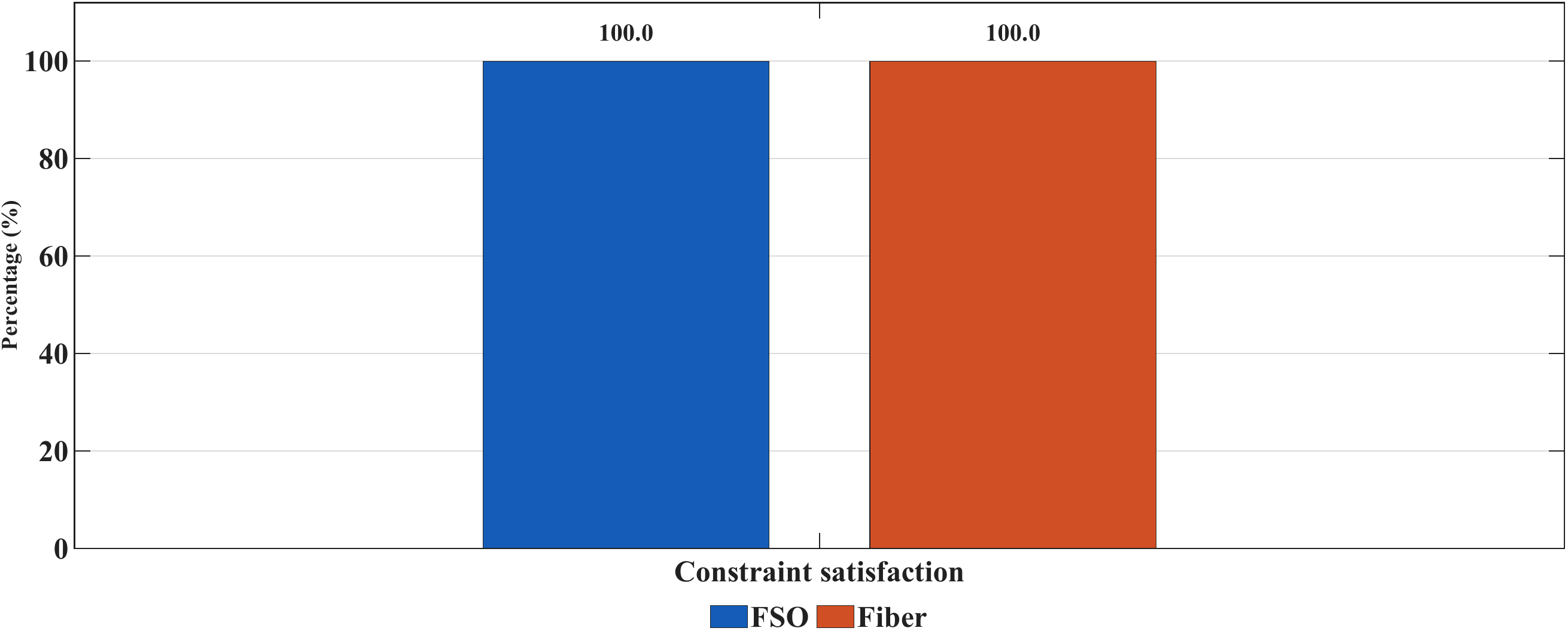}
        \caption{\small Constraint satisfaction of the proposed HC-PPO algorithm for both fiber and FSO channels. }
        \label{fig:constraint}
    \end{figure}

    Fig. \ref{fig:constraint} reports the observed constraint satisfaction on the held-out test instances. The proposed HC-PPO method achieved $100\%$ observed feasibility for both FSO and fiber links, i.e., all evaluated test episodes satisfied constraints $\mathrm{C1}$--$\mathrm{C8}$ under the considered simulation settings. The reported feasibility refers to the complete scheme, i.e., the actor followed by the shield. The actor alone did not guarantee feasibility; whenever its allocation violated the sensing-reachability or reliability conditions, it was replaced by the shield's allocation, and the two components together satisfied the constraints on all evaluated test episodes.
    
\subsection{Realized Terminal Sensing-Target Attainment}
     Constraint $\mathrm{C3}$ ensures that the expected terminal information meets the sensing target, but random preparation and delivery failures can leave the realized information below that target. This experiment quantifies how often the target is reached in individual realizations under the proposed HC-PPO framework. We replayed the trained framework on frozen test instances and sampled each selected probe's preparation and delivery outcome, accumulating the QFI of successful probes according to \eqref{eq:realized_qfi}. Table \ref{tab:realized} reports $\widehat{\Pr}\big(\mathcal{J}_{T+1}^{\mathrm{real}} \geq\mathcal{J}^{\mathrm{target}}\big)$ using $2000$ independent Monte Carlo realizations per test instance and $100$ test instances per training seed, for FSO and fiber links at $\mathcal{J}^{\mathrm{target}}=25$ and $100$. The mean expected terminal information was approximately $1.6$--$12.6$ times the target, while realized information reached the target in approximately $96$--$98\%$ of realizations. Thus, substantial expected-information surplus does not ensure terminal-target attainment in every realization. Constraint $\mathrm{C4}$ separately controls per-slot reliability; the reported terminal success probabilities are empirical properties of the evaluated HC-PPO framework, not guarantees.
    \begin{table}[t]
    \centering
    \caption{Realized terminal-target attainment probability and mean expected-information-to-target ratio. Probability estimates include $95\%$ confidence-interval half-widths across five training seeds, expressed in percentage points (pp).}
    \label{tab:realized}
    \begin{tabular}{lcccc}
    \hline
    Channel & $\mathcal{J}^{\mathrm{target}}$ & $\widehat{\Pr}(\cdot)$ [\%] & 95\% CI [$\pm$ pp] & $\mathbb{E}[\mathcal{J}_{T+1}]/\mathcal{J}^{\mathrm{target}}$ \\
    \hline
    FSO   & 25  & 97.3 & 0.02 & 3.95 \\
    Fiber & 25  & 95.8 & 4.3  & 1.57 \\
    FSO   & 100 & 97.8 & 0.7  & 12.6 \\
    Fiber & 100 & 98.0 & 3.1  & 11.4 \\
    \hline
    \end{tabular}
    \end{table}
    
\subsection{GHZ Probing Versus Separable Probing}

\begin{figure}
    \centering
    \includegraphics[width=0.7\linewidth]{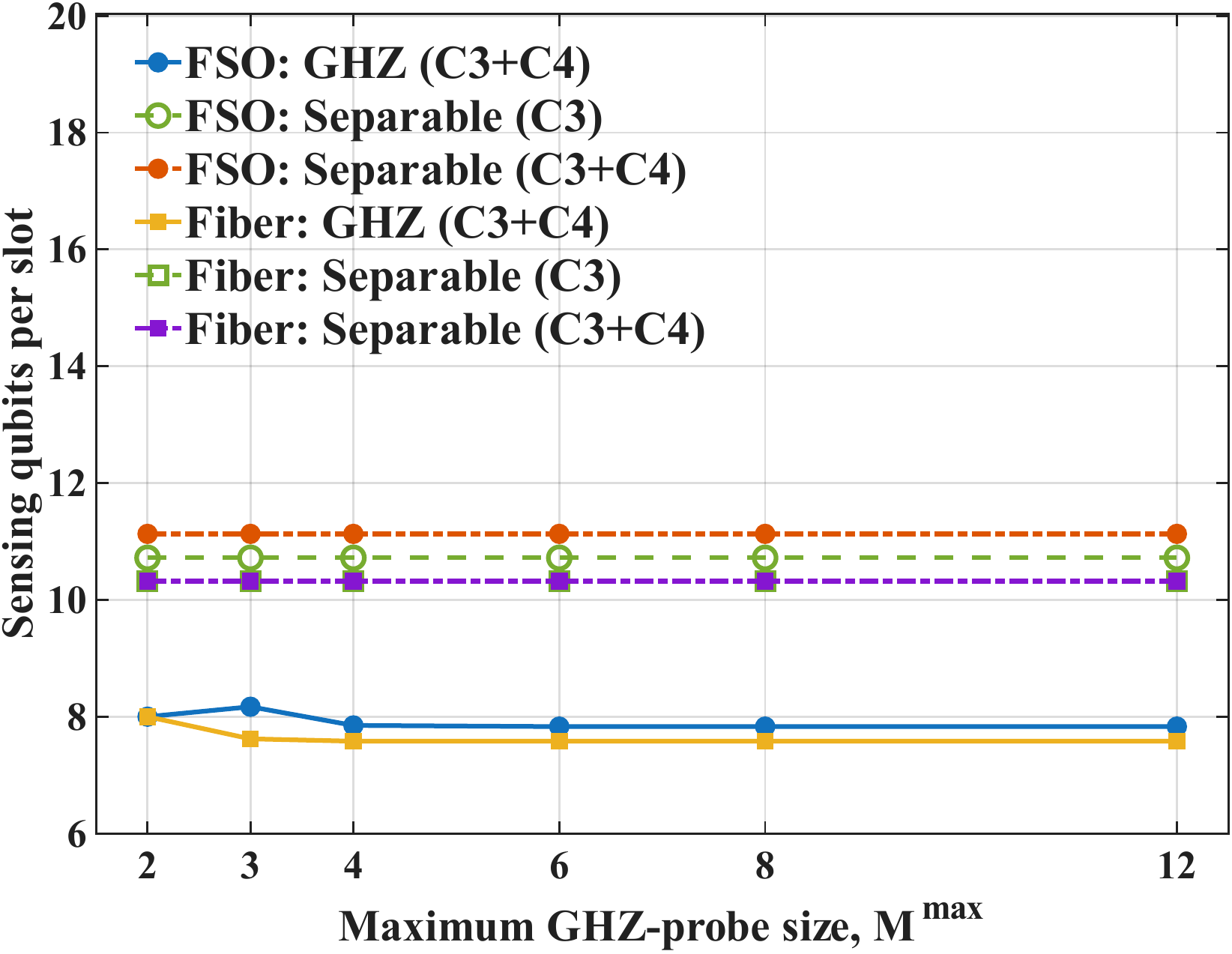}
    \caption{Sensing-resource requirements versus the maximum GHZ-probe size $M^{\max}$ at $J^{\mathrm{target}}=100$  for FSO   and fiber channels.}
    \label{fig:ghz_sep_mmax}
\end{figure}

\begin{figure}
    \centering
    \includegraphics[width=0.775\linewidth]{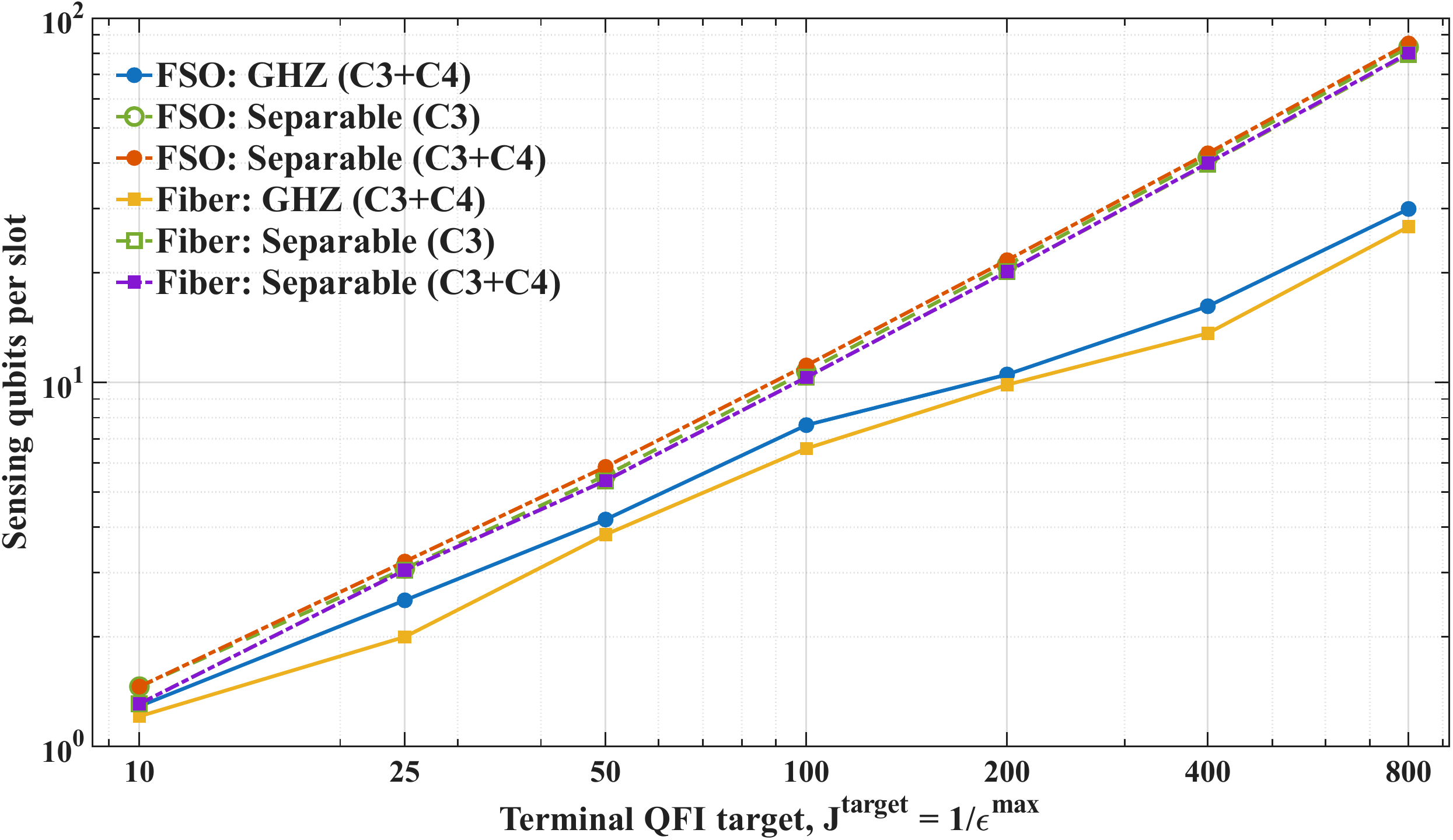}
    \caption{Sensing-resource requirements versus the information target
    $J^{\mathrm{target}}=1/\epsilon^{\max}$ for FSO   and fiber channels.}
    \label{fig:ghz_sep_jtarget}
\end{figure}

    We compared the sensing-resource requirements of GHZ probing and separable probing under the same prescribed sensing target. The separable scheme employed independent, non-entangled single-qubit probes. Two separable variants were considered: one subject to the expected-information requirement $\mathrm{C3}$ only, and another subject to both $\mathrm{C3}$ and the sensing-reliability requirement $\mathrm{C4}$. GHZ probing was evaluated under $\mathrm{C3}$ and $\mathrm{C4}$. The sensing target was specified externally, rather than matched to any surplus expected QFI produced by the GHZ allocation. The terminal expected-information target was $J^{\mathrm{target}}$. In these resource-requirement experiments, a per-slot threshold of $J^{\mathrm{target}}/T$, with $T=8$, was applied in both the expected-information and sensing-reliability tests. Figs. \ref{fig:ghz_sep_mmax} and \ref{fig:ghz_sep_jtarget} show the mean sensing-qubit consumption per slot for FSO and fiber channels. For each scheme, the mean was calculated over the instances for which its allocation procedure returned an allocation satisfying the imposed sensing constraints.

    Fig. \ref{fig:ghz_sep_mmax} illustrates the effect of the maximum GHZ-probe size $M^{\max}$ at $J^{\mathrm{target}}=100$. The separable requirements did not depend on this GHZ-specific cap: the $\mathrm{C3}$+$\mathrm{C4}$ variant used an average of $11.13$ qubits per slot under FSO and $10.32$ under fiber. The GHZ resource consumption leveled off at modest probe-size caps, reaching $7.83$ qubits per slot for FSO at the tested values $M^{\max}\geq 6$ and $7.58$ for fiber at the tested values $M^{\max}\geq 4$. These values correspond to reductions of $29.6\%$ and $26.6\%$, respectively, relative to the separable $\mathrm{C3}$+$\mathrm{C4}$ mean. Further increases in the cap provided no additional reduction in the reported means over the tested range $M^{\max}\leq12$. Thus, modest probe-size caps were sufficient to obtain the observed resource savings in this configuration.

    Fig. \ref{fig:ghz_sep_jtarget} examines sensing-information targets from $10$ to $800$, where a larger $J^{\mathrm{target}}=1/\epsilon^{\max}$ corresponds to a more stringent sensing-accuracy requirement. The separable $\mathrm{C3}$+$\mathrm{C4}$ resource requirement grew approximately in proportion to the target, increasing from $1.46$ to $85.26$ qubits per slot for FSO and from $1.31$ to $80.25$ for fiber. At target $J^{\mathrm{target}}=25$, the separable $\mathrm{C3}$+$\mathrm{C4}$ and GHZ means were $3.22$ and $2.52$ qubits per slot for FSO, and $3.05$ and $2.00$ for fiber. 
    Substantially larger savings were observed at the highest tested target. At $J^{\mathrm{target}}=800$, adaptive GHZ probing used approximately $29.94$ qubits per slot for FSO and $26.76$ for fiber, compared with $85.26$ and $80.25$ for the separable $\mathrm{C3}$+$\mathrm{C4}$ variant. These values corresponded to reductions of approximately $64.9\%$ and $66.7\%$, respectively, in the reported mean sensing-resource consumption.
    Imposing $\mathrm{C4}$ on the separable scheme added only a modest sensing-resource cost in the evaluated settings. For FSO, the $\mathrm{C3}$+$\mathrm{C4}$ mean exceeded the $\mathrm{C3}$-only mean by $4.5\%$ at $J^{\mathrm{target}}=25$ ($3.22$ versus $3.08$) and by $2.4\%$ at $J^{\mathrm{target}}=800$ ($85.26$ versus $83.30$). For fiber, the two variants coincided at the tested targets up to $200$ and differed by less than $1\%$ at the larger tested targets. The evaluated GHZ allocations reduced mean sensing-resource consumption by approximately $65$-$67\%$ at the highest tested target, while the absolute savings remained small at the  default target.

    \begin{figure}
        \centering
        \subfloat[]{
            \includegraphics[width=0.775\linewidth]{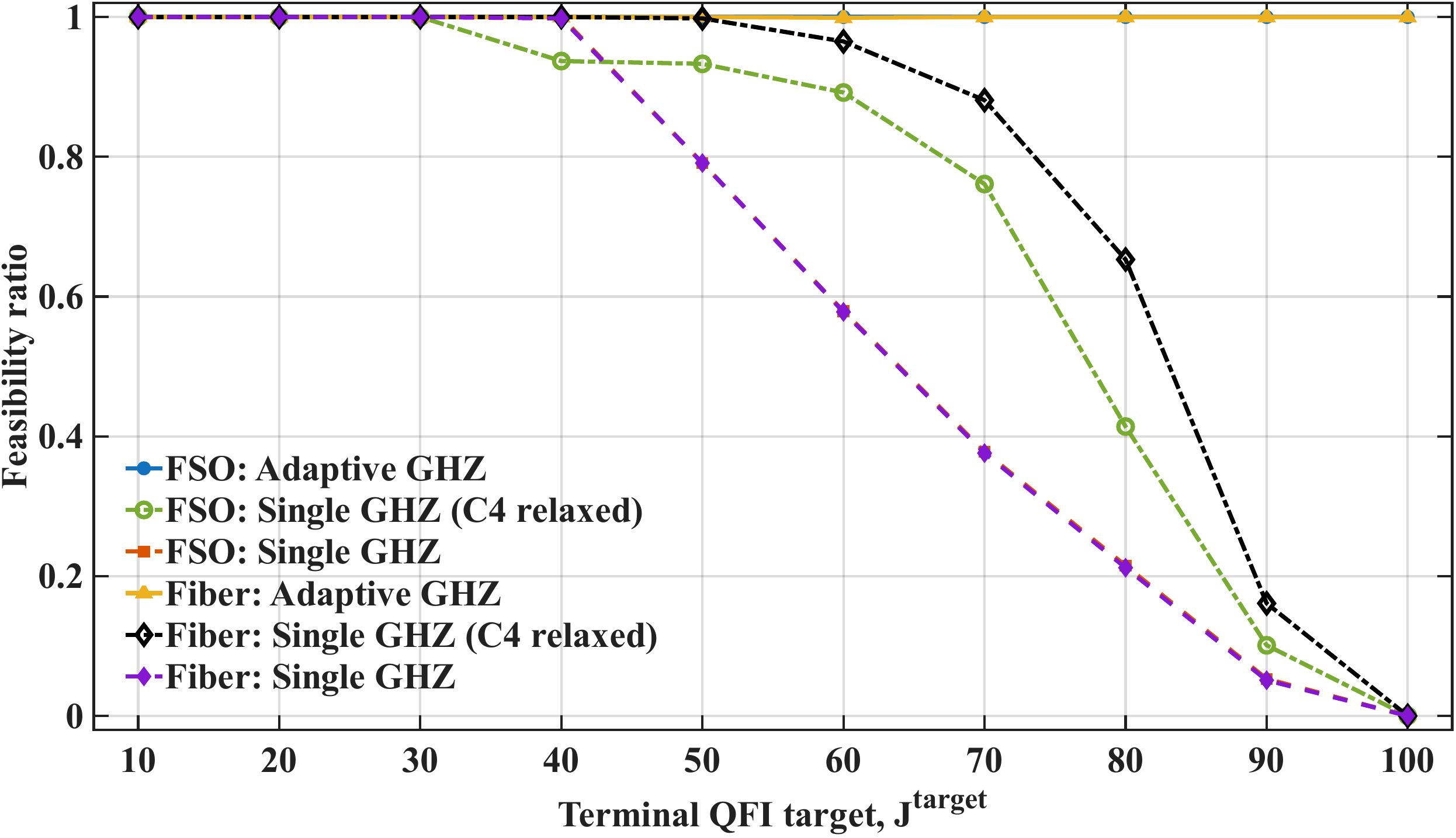}
            \label{fig:adaptive_single_FR_Jtarget}
        }
        
        \subfloat[]{
            \includegraphics[width=0.775\linewidth]{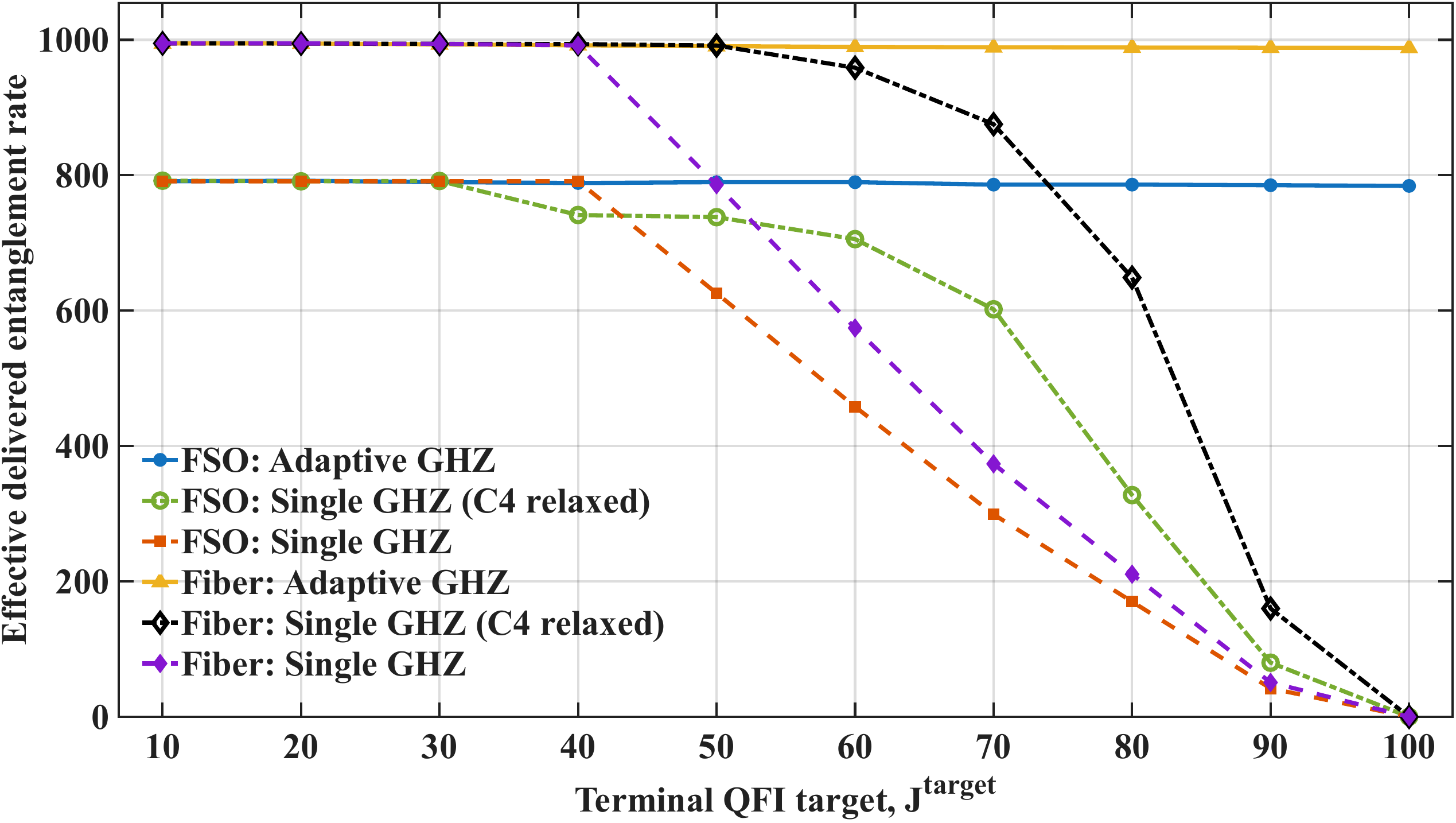}
            \label{fig:adaptive_single_ER_Jtarget}
        }
        
        \caption{\small Performance comparison of adaptive GHZ partitioning and         the fixed single-GHZ structure vs. the terminal QFI target $\mathcal J^{\mathrm{target}}$: (a)  feasibility ratio and
        (b) effective delivered entanglement rate.}
        \label{fig:Adaptive_Single_Jtarget}
    \end{figure}
    
\subsection{Adaptive GHZ Partitioning Versus Single-GHZ Sensing}
    Motivated by sensing architectures that employ a single multipartite  GHZ probe, we  next investigated the benefit of adapting the GHZ partitioning structure by comparing the proposed adaptive multi-GHZ scheme with a single-GHZ baseline \cite{single-ghz,GHZ-partitioning}. In the baseline, at most one GHZ probe could be active in each slot, i.e., $G_t\leq 1$, while its size and sensing-node composition remained dynamically adjustable. By contrast, the proposed scheme can adapt both the number and sizes of simultaneously active GHZ probes according to the sensing requirement, channel conditions, and available resources.
    To jointly account for communication performance and the ability to satisfy the system constraints, we report the feasibility-weighted effective delivered entanglement rate as $R_{\mathrm{eff}}    =    P_{\mathrm{feas}}\, \bar{R}_{\mathrm{feas}}$, where $P_{\mathrm{feas}}$ denotes the fraction of feasible test realizations and $\bar{R}_{\mathrm{feas}}$ is the average delivered entanglement rate conditioned on feasibility. Hence, $R_{\mathrm{eff}}$ is equivalent to assigning zero usable communication service to infeasible realizations. In the following Figs. \ref{fig:Adaptive_Single_Jtarget} and \ref{fig:Adaptive_Single_Pfus}, in addition to the adaptive and single-GHZ schemes evaluated under the original constraints, we included a single-GHZ variant with $\mathrm{C4}$ relaxed to isolate the effect of the sensing-reliability requirement. This variant serves as a diagnostic comparison and was evaluated under a less restrictive constraint set. 

    Fig.~\ref{fig:Adaptive_Single_Jtarget} investigates the effect of the terminal sensing requirement $\mathcal J^{\rm target}$. 
    As shown in Fig.~\ref{fig:adaptive_single_FR_Jtarget}, the adaptive scheme achieved an observed feasibility ratio of at least $0.999$ across the tested range $\mathcal J^{\rm target}\in[10,100]$ for both FSO and fiber channels. By contrast, the feasibility ratio of the single-GHZ baseline decreased as the sensing target increased. With $\mathrm{C4}$ enforcement, it fell to approximately $0.79$ at $\mathcal J^{\rm target}=50$ and $0.21$ at $\mathcal J^{\rm target}=80$ for both channels. Relaxing $\mathrm{C4}$ substantially delayed this decline: at $\mathcal J^{\rm target}=80$, the corresponding ratios amounted to approximately $0.41$ for FSO and $0.65$ for fiber. However, both single-GHZ variants reached zero observed feasibility at $\mathcal J^{\rm target}=100$. Thus, the reliability constraint contributed substantially to the degradation, but did not fully explain the single-GHZ limitation at the highest sensing targets. Under the imposed probe-size cap $M^{\max}=3$, the information available from a single probe was insufficient at the highest target, even with $\mathrm{C4}$ relaxed. The observed limitation therefore reflects both the reliability requirement and the considered probe-size restriction.
    
    The corresponding communication performance is shown in Fig.~\ref{fig:adaptive_single_ER_Jtarget}. The adaptive scheme maintained a nearly constant effective delivered entanglement rate, approximately $790$ for FSO and $990$ for fiber, with only a slight decrease as the sensing target increased. The effective rates of the single-GHZ variants instead decreased in conjunction with their feasibility ratios and reached zero at the largest tested target. Relaxing $\mathrm{C4}$ preserved a higher effective rate at intermediate targets, particularly under fiber. 
    Adaptive partitioning allows the sensing resources to be distributed among multiple probes whose sizes and node assignments can be adjusted to the operating conditions. The aggregate information can then be supplied by several successful probes, reducing reliance on the successful preparation and delivery of one larger probe. In the evaluated configurations, this flexibility supported higher sensing targets while maintaining the communication service under the original constraint set. 


    \begin{figure}[!t]
        \centering
    
        \subfloat[]{
            \includegraphics[width=0.775\linewidth]
            {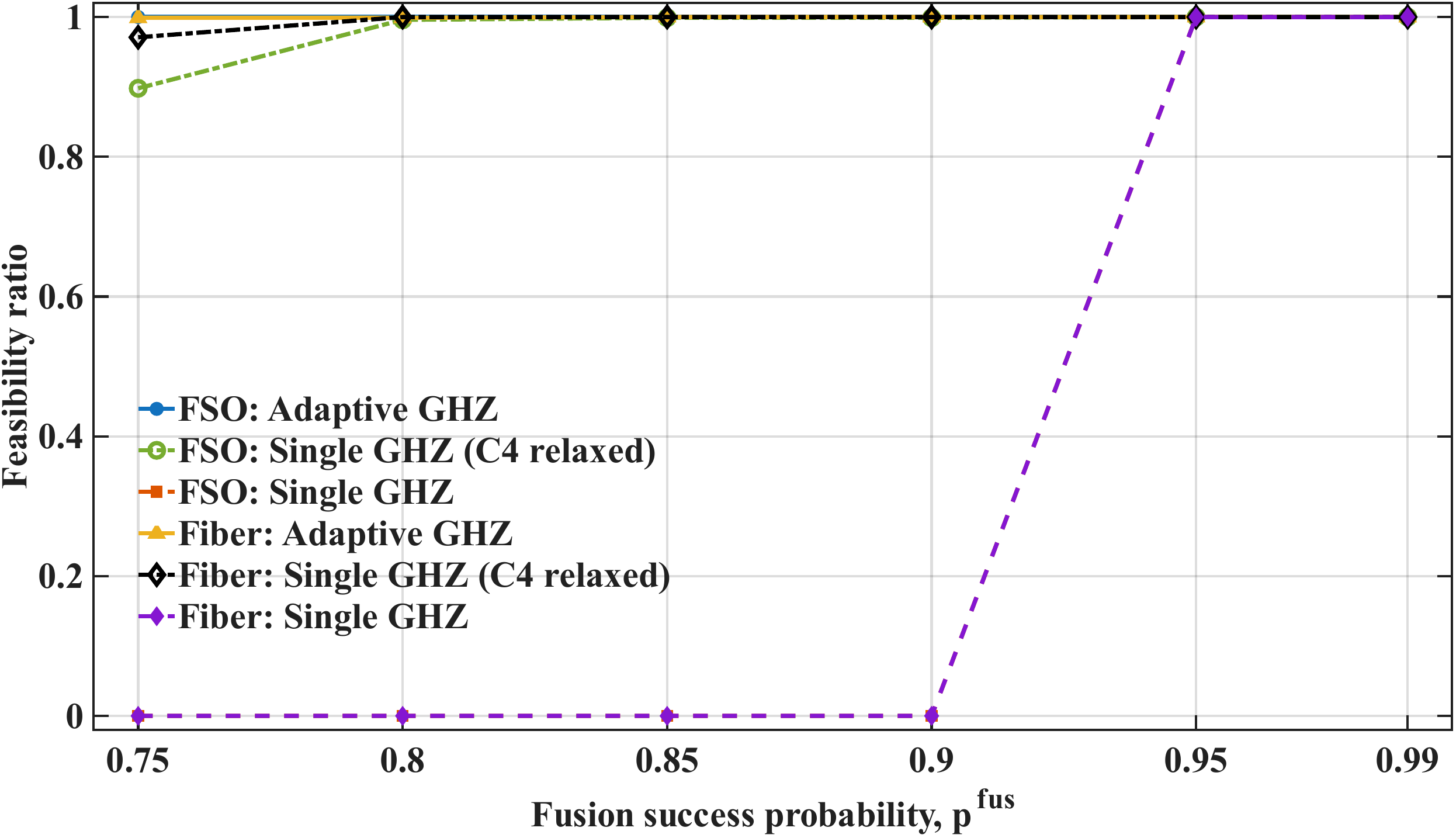}
            \label{fig:adaptive_single_FR_Pfus}
        }
    
        \subfloat[]{
            \includegraphics[width=0.775\linewidth]
            {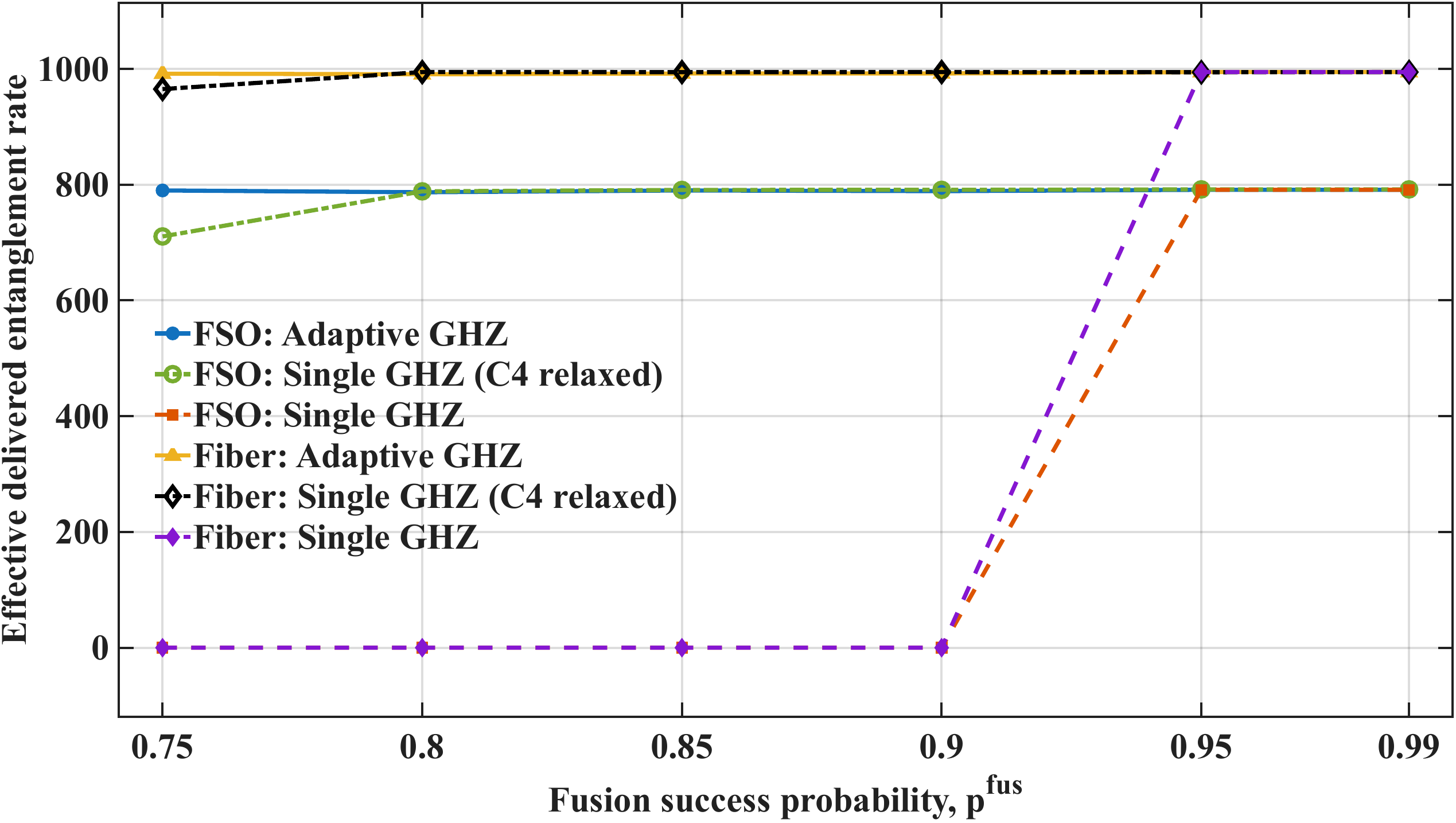}
            \label{fig:adaptive_single_ER_Pfus}
        }
    
        \caption{\small Performance comparison of adaptive GHZ partitioning and
        the fixed single-GHZ structure vs. the fusion success probability
        $p_{\mathrm{fus}}$: (a) feasibility ratio  and
        (b) effective delivered entanglement rate.}
        \label{fig:Adaptive_Single_Pfus}
    \end{figure}
   The impact of imperfect GHZ-state preparation is depicted in Fig. \ref{fig:Adaptive_Single_Pfus} by varying the fusion success probability $p_{\mathrm{fus}}$. 
    As shown in Fig.~\ref{fig:adaptive_single_FR_Pfus}, the adaptive scheme maintained an observed feasibility ratio close to unity across the tested fusion probabilities for both FSO and fiber channels. With $\mathrm{C4}$ enforced, the single-GHZ baseline had zero observed feasibility at the tested values $p_{\mathrm{fus}}\leq0.9$ and reached unity at $p_{\mathrm{fus}}=0.95$ and $0.99$. Relaxing $\mathrm{C4}$ substantially restored its feasibility, demonstrating that the reliability requirement was the principal cause of the single-GHZ collapse in this experiment. This behavior followed from the preparation-and-delivery success probability of a single GHZ probe. A probe containing $m$ distributed sensing qubits and one QBS reference qubit requires $m-1$ successful fusion operations, giving a preparation-success factor $p_{\mathrm{fus}}^{m-1}$, which is further reduced by transmission losses. When the per-slot information threshold cannot be met by a single-qubit probe, the single-GHZ scheme must use a larger probe. If its preparation-and-delivery success probability falls below $1-\delta_s$, it cannot satisfy $\mathrm{C4}$, even when its expected information is sufficient to satisfy the information requirement. Adaptive partitioning can distribute sensing resources among several probes, allowing the aggregate information  from successful probes to meet the threshold despite failures of other probes. 
    
    Fig. \ref{fig:adaptive_single_ER_Pfus} further shows that the reported communication rates of the adaptive scheme and the $\mathrm{C4}$-relaxed single-GHZ variant were close throughout the tested range, with the relaxed variant sometimes attaining a slightly higher rate. At $p_{\mathrm{fus}}=0.95$ and $0.99$, the original single-GHZ baseline also achieved a similar rate. Thus, the main advantage demonstrated by this experiment is the adaptive scheme's ability to accommodate the prescribed sensing reliability over a wider range of fusion probabilities, instead of a uniform communication-rate improvement over single-GHZ probing. 
\subsection{Joint Communication--Sensing Co-Design Versus Fixed Resource Splitting}
    \begin{figure}[!t]
        \centering
        \includegraphics[width=0.775\linewidth]{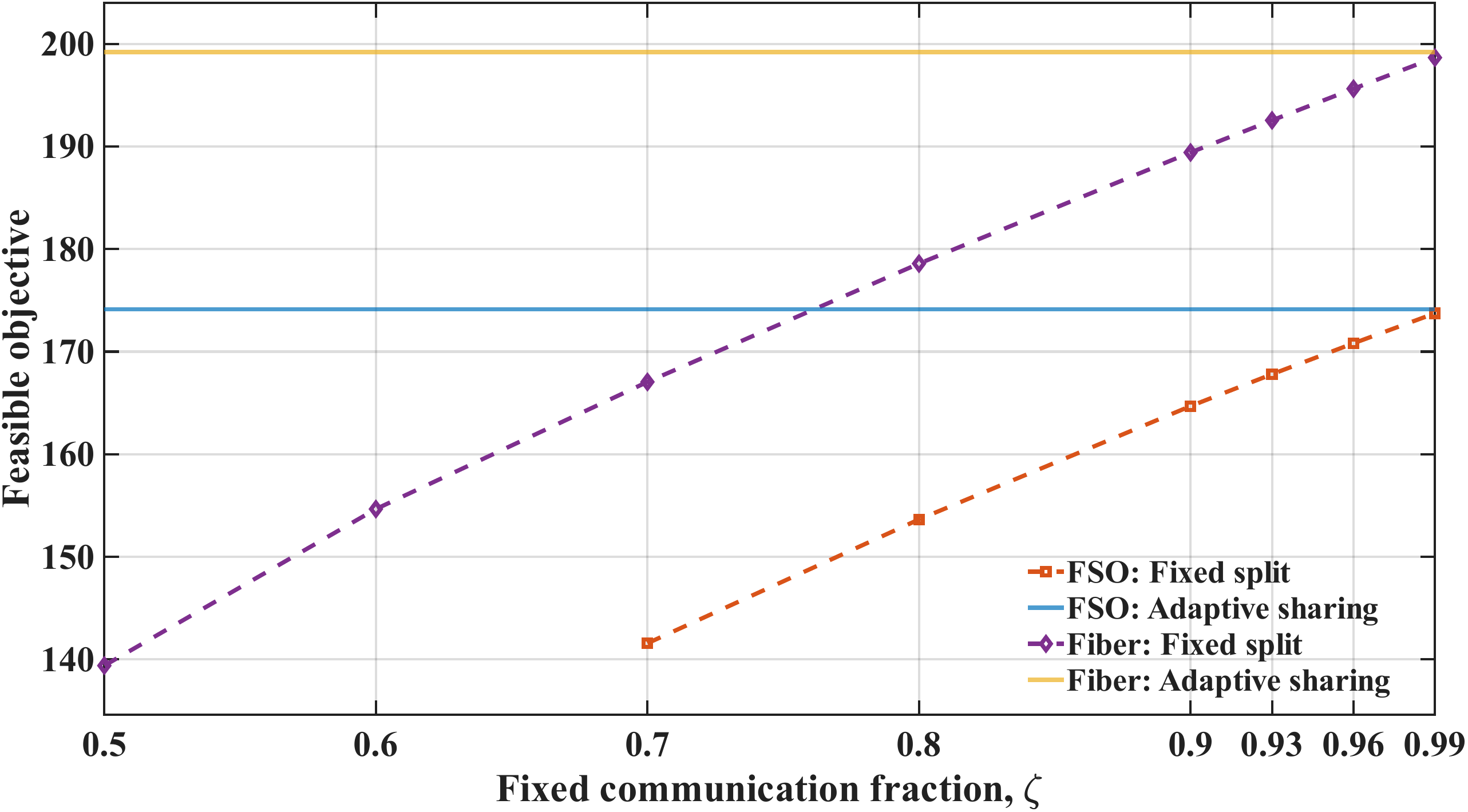}
        \caption{\small Objective value vs. the fixed communication fraction     $\zeta$ for communication--sensing co-design and fixed resource     splitting over FSO and fiber links, with $N^{\max}=1000$ and $p_{\mathrm{fus}}=0.9$.}
        \label{fig:adaptive_fixed_portion}
    \end{figure}
    
    \begin{figure}[!t]
        \centering
        \includegraphics[width=0.775\linewidth]{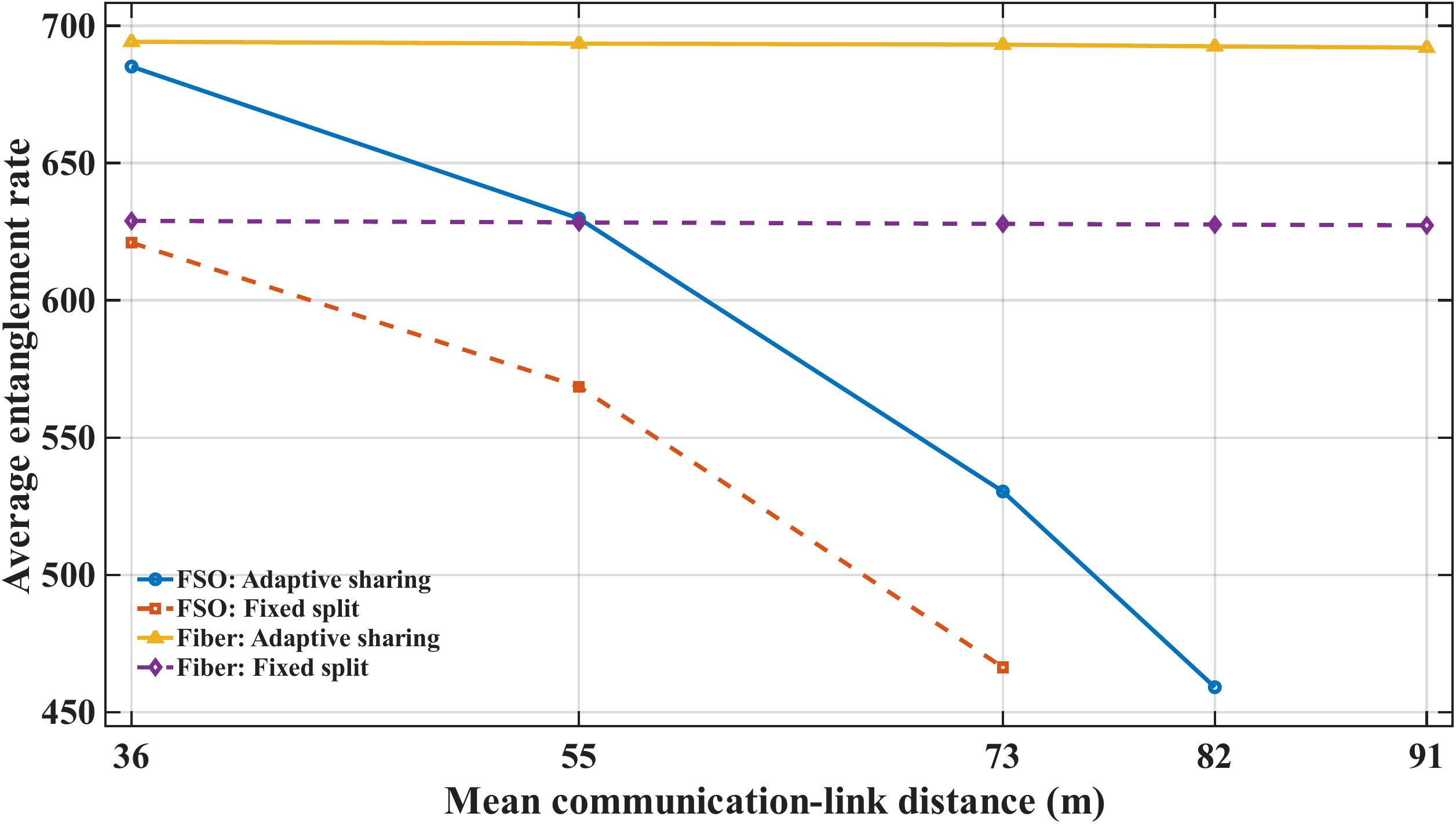}
        \caption{\small Average  entanglement rate vs. the mean communication-link distance for communication--sensing co-design and fixed resource splitting over FSO and fiber links, with $\zeta=0.9$, $p_{\mathrm{fus}}=0.9$, and $N^{\max}=700$.}
        \label{fig:adaptive_fixed_distance}
    \end{figure}
    In this section, we evaluated the benefit of jointly adapting the communication and sensing resource allocation. The proposed framework dynamically determined the amount of the shared Bell-pair budget assigned to sensing in each slot, with the residual resources available for communication. As a benchmark, we considered a fixed resource-splitting strategy where $\lfloor\zeta N^{\max}\rfloor$ Bell pairs were reserved for communication, while the remaining budget was reserved for sensing in every slot. Thus, unlike the proposed adaptive scheme, the fixed-split baseline could not adjust the communication--sensing balance according to the instantaneous resource requirements and channel conditions.

    Fig. \ref{fig:adaptive_fixed_portion} compares the mean feasible objective achieved by the proposed co-design strategy with that obtained under different fixed values of $\zeta$ for $N^{\max}=1000$. Since the adaptive scheme does not use a predetermined splitting factor, its performance is shown as a horizontal reference for each channel configuration. By contrast, the performance of the fixed-split scheme strongly depends on $\zeta$. Over the feasible fixed-split configurations, increasing $\zeta$ generally improved the communication objective, as a larger fraction of the entanglement-generation budget was reserved for communication. However, this benefit could be realized only while the remaining sensing budget was sufficient to satisfy the sensing constraints.  
    For the FSO configuration, the fixed-split objective approached the adaptive result only for strongly communication-oriented allocations, whereas smaller communication fractions incurred a considerable objective loss. A similar trend was observed for fiber links, for which the fixed-split performance approached but did not exceed the adaptive-sharing result over the considered range. At $\zeta=0.99$, fixed splitting nearly matched adaptive sharing for both channels. Thus, the benefit of adaptive sharing was small relative to a suitably chosen fixed split in this setting, whereas smaller communication fractions incurred substantial losses by reserving more resources for sensing.

    To examine whether this advantage would persist under changing propagation  conditions, Fig. \ref{fig:adaptive_fixed_distance} shows the average delivered entanglement rate as a function of the mean communication-link distance. The fixed-split benchmark used $\zeta=0.9$ throughout the distance sweep, whereas the proposed scheme continued to adapt the communication--sensing allocation. For FSO links, the delivered entanglement rate noticeably decreased with increasing communication-link distance for both schemes. This behavior is consistent with the stronger distance-dependent losses of the considered FSO channel, including geometric spreading and pointing effects. Yet, the adaptive sharing strategy achieved a higher delivered entanglement rate than the fixed-split baseline over their common operating region. 
    For fiber links, the delivered entanglement rate remained comparatively stable over the considered distance range, reflecting the weaker distance dependence under the adopted parameters. However, adaptive sharing maintained a performance advantage over the fixed resource split across the evaluated distances. Thus, even when propagation degradation was moderate, dynamically sharing the Bell-pair budget enabled more effective use of the common communication--sensing resources than a predetermined static partition.

    \begin{figure}[!t]
        \centering
        \includegraphics[width=0.775\linewidth]{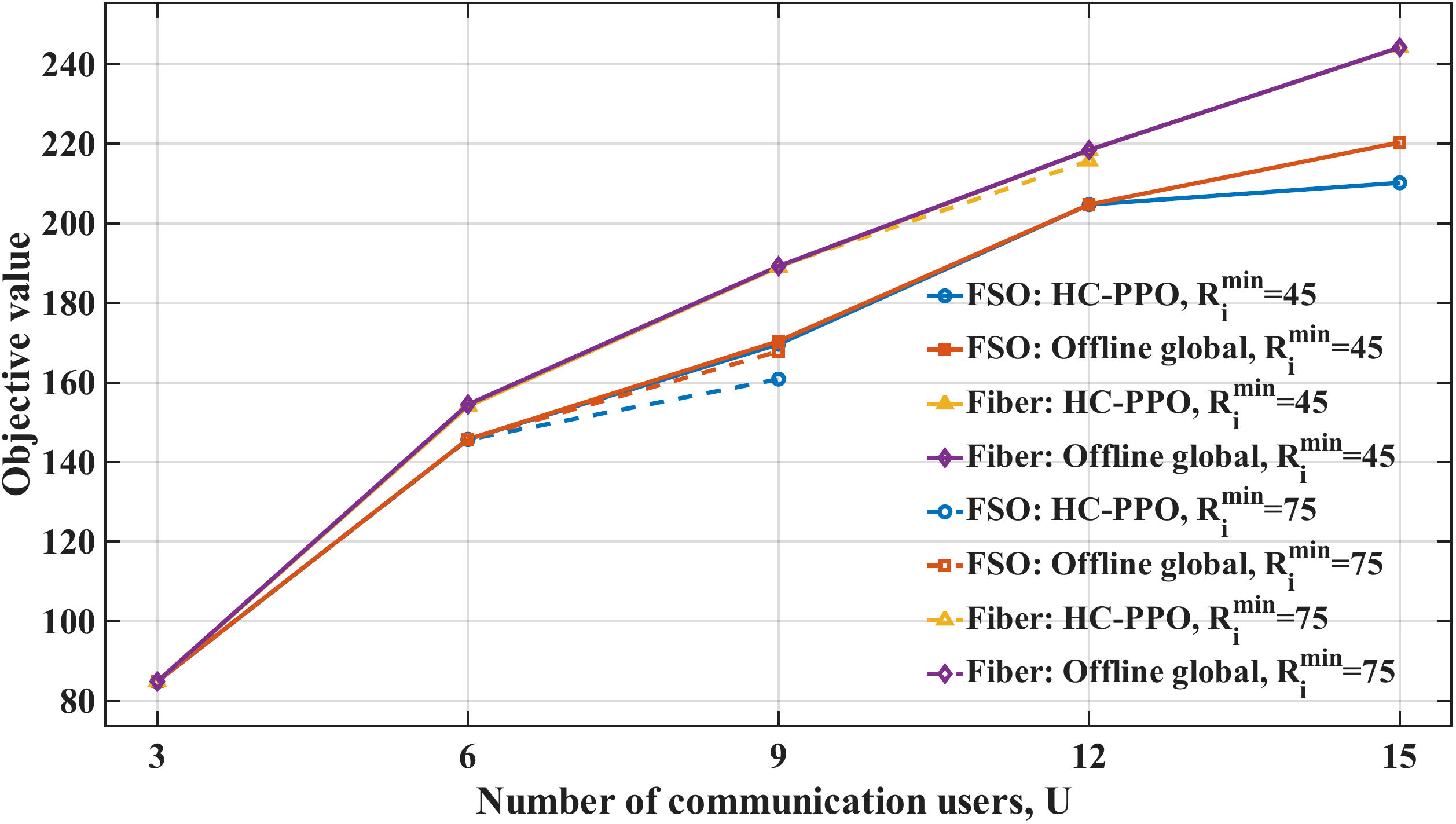}
        \caption{\small Objective value vs. the number of communication users     $U$ for HC-PPO and the offline global benchmark over FSO and fiber links     under $R_i^{\min}=45$ and $R_i^{\min}=75$. }
        \label{fig:optimality_gap}
    \end{figure}
\subsection{Optimality Gap}
    We evaluated the solution quality of the proposed HC-PPO framework by comparing its achieved objective with an offline global optimization benchmark in this section. Since obtaining a globally optimal solution becomes computationally demanding for large problem instances, this comparison was performed for a moderate number of communication users, $U\in\{3,6,9,12,15\}$. Feasible discrete sensing configurations were enumerated offline, after which the resulting full-horizon mixed-integer optimization problem was solved using the Gurobi Optimizer~\cite{gurobi} with a time limit of $300$\,s per instance and a relative optimality gap tolerance of $0.1\%$. Only instances certified optimal within the prescribed $0.1\%$ relative tolerance over the enumerated sensing configurations were included in the comparison.     We considered both FSO and fiber links and two minimum communication-rate requirements, namely, $R_i^{\min}=45$ and $R_i^{\min}=75$, in order to examine the behavior of the proposed method under different QoS requirements. 
    
    As shown in Fig. \ref{fig:optimality_gap}, for $\bar R_i^{\min}=45$, the maximum reported objective gap between HC-PPO and the certified offline benchmark was approximately $4.611\%$, occurring for FSO at $U=15$. In particular, this indicates that the HC-PPO solution can recover solutions very close to the globally optimized resource allocation. The largest deviation occurred for the FSO configuration at $U=15$, where the optimality gap was approximately $4.611\%$.  For $ R^{\min}=75$, the FSO configuration became infeasible for $U\geq 12$, while the fiber configuration became infeasible at $U=15$; accordingly, no objective values are reported for these operating points. 
    Increasing the minimum-rate requirement to $R_i^{\min}=75$ reduced the feasible operating region. For FSO links, feasible solutions were obtained up to $U=9$, whereas the fiber configuration remained feasible up to $U=12$. Importantly, HC-PPO continued to closely track the offline global benchmark  within the feasible region under the more stringent QoS requirement. The maximum observed optimality gap for $R_i^{\min}=75$ amounted to approximately $4.148\%$, occurring for the FSO case, while the gap for the feasible fiber configurations remained smaller. Therefore, increasing the communication requirement primarily reduced the set of feasible network sizes, rather than caused a systematic loss of optimality in the proposed policy.  
    At $U=15$, HC-PPO, including the shield, produced a complete allocation in approximately $400$\,ms per slot ($3.2$\,s per horizon) for FSO and approximately $231$\,ms per slot ($1.8$\,s per horizon) for fiber, on a single CPU core.  
    These results indicate that HC-PPO provides near-optimal resource-allocation decisions over the feasible operating region while avoiding the need to repeatedly solve the offline global optimization problem during online operation. These results provide empirical evidence that the proposed HC-PPO achieves a favorable compromise between solution quality and online applicability.

\section{Conclusion}\label{conclusion}
    In this paper, we investigated joint entanglement-resource allocation for quantum integrated sensing and communication networks where communication and distributed sensing share a common supply of elementary Bell pairs. We developed an adaptive multi-GHZ sensing framework that dynamically determines the probe sizes and sensing-node assignments while accounting for probabilistic fusion, optical transmission loss, and quantum-memory decoherence. The resulting finite-horizon resource-allocation problem captures jointly communication QoS, sensing accuracy and reliability, and finite generation, processing, memory, and fusion resources. To address its mixed discrete--continuous structure, we proposed an HC-PPO framework combining constrained-PPO sensing allocation with analytical QoS-aware $\rho$-fair communication allocation.  
    The numerical results over both FSO and fiber links demonstrated that adaptive multi-GHZ probing provides improved resource efficiency and robustness compared with single-GHZ sensing and separable probing. Moreover, dynamically sharing the available Bell-pair resources between communication and sensing was found to outperform fixed resource splitting across the considered operating conditions. The proposed framework satisfied all constraints on every held-out test instance in the convergence experiment; this empirical result does not imply a joint feasibility guarantee from the masks and shield. These results highlight the benefit of jointly adapting multipartite-entanglement structures and shared quantum resources in Q-ISAC networks. Future work can extend the proposed framework to time-varying channel realizations, fidelity-aware multipartite-state construction, and more detailed quantum hardware imperfections.

	\bibliographystyle{IEEEtran}
    \balance
    \bibliography{Refs}

@ARTICLE{channel-modeling,
  author={Najafi, Marzieh and others},
  journal={IEEE Transactions on Communications}, 
  title={Statistical Modeling of the {FSO} Fronthaul Channel for {UAV}-Based Communications}, 
  year={2020},
  volume={68},
  number={6},
  pages={3720-3736}
}

@ARTICLE{chehimi-ris,
  author={Chehimi, Mahdi and others},
  journal={IEEE Transactions on Wireless Communications}, 
  title={Reconfigurable Intelligent Surface ({RIS})-Assisted Entanglement Distribution in {FSO} Quantum Networks}, 
  year={2025},
  volume={24},
  number={4},
  pages={3132-3148}
}

@ARTICLE{pointing-error,
  author={Jung, Kug-Jin and Nam, Sung Sik and Alouini, Mohamed-Slim and Ko, Young-Chai},
  journal={IEEE Transactions on Communications}, 
  title={Unified Finite Series Approximation of {FSO} Performance Over Strong Turbulence Combined With Various Pointing Error Conditions}, 
  year={2020},
  volume={68},
  number={10},
  pages={6413-6425}
}

@article{GHZ-fusion,
  title={Protocols for creating and distilling multipartite {GHZ} states with {Bell} pairs},
  author={de Bone, S{\'e}bastian and Ouyang, Runsheng and Goodenough, Kenneth and Elkouss, David},
  journal={IEEE Transactions on Quantum Engineering},
  volume={1},
  pages={1--10},
  year={2020},
  publisher={IEEE}
}

@article{global-phase,
  title={Optimal and secure measurement protocols for quantum sensor networks},
  author={Eldredge, Zachary and others},
  journal={Physical Review A},
  volume={97},
  number={4},
  pages={042337},
  year={2018},
  publisher={APS}
}

@article{QFI-constant,
  title={Multiparameter estimation in networked quantum sensors},
  author={Proctor, Timothy J and Knott, Paul A and Dunningham, Jacob A},
  journal={Physical review letters},
  volume={120},
  number={8},
  pages={080501},
  year={2018},
  publisher={APS}
}

@ARTICLE{fiber-link,
  author={Dai, Wenhan and Peng, Tianyi and Win, Moe Z.},
  journal={IEEE Journal on Selected Areas in Communications}, 
  title={Optimal Remote Entanglement Distribution}, 
  year={2020},
  volume={38},
  number={3},
  pages={540-556}
  }

@article{bosonic,
  title={Fundamental limits of repeaterless quantum communications},
  author={Pirandola, Stefano and Laurenza, Riccardo and Ottaviani, Carlo and Banchi, Leonardo},
  journal={Nature communications},
  volume={8},
  number={1},
  pages={15043},
  year={2017},
  publisher={Nature Publishing Group UK London}
}

@article{Shiva-tutorial,
  title={A Tutorial on Resource Management in Quantum Networks: Basics, State-of-the-Art, and Open Challenges},
  author={Taskooh, Shiva Kazemi and Hossain, Ekram and Kaddoum, Georges},
  year={2026}
}

@article{GHZ-partitioning,
  title={Enhancing noisy quantum sensing by {GHZ} state partitioning},
  author={Zang, Allen and others},
  journal={arXiv preprint arXiv:2507.02829},
  year={2025},
  url={https://arxiv.org/pdf/2507.02829}
}

@article{QISAC-entanglement,
  title={Quantum integrated sensing and communication via entanglement},
  author={Liu, Yu-Chen and others},
  journal={Physical Review Applied},
  volume={22},
  number={3},
  pages={034051},
  year={2024},
  publisher={APS}
}

@ARTICLE{variational,
  author={Shohibul Ulum, Muhammad and others},
  journal={IEEE Journal on Selected Areas in Communications}, 
  title={Variational Anonymous Quantum Sensing}, 
  year={2024},
  volume={42},
  number={9},
  pages={2275-2291}
  }

@article{qisac-monemi,
  title={Quantum Integrated Sensing and Communication: Foundations, Applications, and Future Directions},
  author={Monemi, Mehdi and others},
  journal={IEEE Communications Magazine},
  year={2026},
  publisher={IEEE}
}

@article{avis-analysis,
  title={Analysis of multipartite entanglement distribution using a central quantum-network node},
  author={Avis, Guus and Rozpedek, Filip and Wehner, Stephanie},
  journal={Physical Review A},
  volume={107},
  number={1},
  pages={012609},
  year={2023},
  publisher={APS}
}

@inproceedings{wang-joint,
  title={Joint quantum communication and sensing},
  author={Wang, Shi-Yuan and Erdo{\u{g}}an, Tuna and Pereg, Uzi and Bloch, Matthieu R},
  booktitle={2022 IEEE Information Theory Workshop (ITW)},
  pages={506--511},
  year={2022},
  organization={IEEE}
}

@article{qisac-variational,
  title={Variational Quantum Integrated Sensing and Communication},
  author={Nikoloska, Ivana and Simeone, Osvaldo},
  journal={arXiv preprint arXiv:2511.16597},
  year={2025}
}

@article{isac-radar,
  title={Quantum Radar for {ISAC}: Sum-Rate Optimization},
  author={Alsaui, Abdulmohsen and others},
  journal={IEEE Transactions on Communications},
  year={2026},
  publisher={IEEE}
}

@article{quantum-advantage,
  title={Quantum advantage in distributed sensing with noisy quantum networks},
  author={Zang, Allen and others},
  journal={Physical Review Research},
  volume={8},
  number={3},
  pages={033040},
  year={2026},
  publisher={APS}
}

@article{single-ghz,
  title={Distributed multi-parameter quantum metrology with a superconducting quantum network},
  author={Zhang, Jiajian and others},
  journal={Nature Communications},
  volume={17},
  number={1},
  pages={1825},
  year={2026},
  publisher={Nature Publishing Group UK London}
}

@misc{gurobi,
  author = {{Gurobi Optimization, LLC}},
  title = {Gurobi Optimizer Reference Manual},
  year = {2026},
  url = {https://www.gurobi.com}
}

@article{masking,
  title={A closer look at invalid action masking in policy gradient algorithms},
  author={Huang, Shengyi and Onta{\~n}{\'o}n, Santiago},
  journal={arXiv preprint arXiv:2006.14171},
  year={2020}
}

@article{sequential-action,
  title={Neural combinatorial optimization with reinforcement learning},
  author={Bello, Irwan and others},
  journal={arXiv preprint arXiv:1611.09940},
  year={2016}
}

@inproceedings{shield,
  title={Safe reinforcement learning via shielding},
  author={Alshiekh, Mohammed and others},
  booktitle={Proceedings of the AAAI conference on artificial intelligence},
  volume={32},
  number={1},
  year={2018}
}

@article{CPPO,
  title={Model-based safe deep reinforcement learning via a constrained proximal policy optimization algorithm},
  author={Jayant, Ashish K and Bhatnagar, Shalabh},
  journal={Advances in Neural Information Processing Systems},
  volume={35},
  pages={24432--24445},
  year={2022}
}
    \vfill

\end{document}